\documentclass[aps,pre,epsf,superscriptaddress,amsmath,amssymb,amsfonts,twocolumn,showpacs]{revtex4-1}

\usepackage{graphicx}
\usepackage{epsfig}
\usepackage{dcolumn}
\usepackage{bm}
\usepackage{braket}
\usepackage{amsmath}
\usepackage{mathtools}
\usepackage{color}
\usepackage{hyperref}
\newcommand{\abs}[1]{\left| #1 \right|} 

\usepackage{hyperref}
\usepackage{multirow}

\usepackage[normalem]{ulem} 

\begin{document}
\title{Many-body quantum dynamics and induced correlations\\ of Bose polarons}

\author{S. I. Mistakidis}
\affiliation{Center for Optical Quantum Technologies, Department of Physics, University of Hamburg, 
Luruper Chaussee 149, 22761 Hamburg Germany}
\author{G.M. Koutentakis}
\affiliation{Center for Optical Quantum Technologies, Department of Physics, University of Hamburg, 
Luruper Chaussee 149, 22761 Hamburg Germany} \affiliation{The Hamburg Centre for Ultrafast Imaging,
Universit\"{a}t Hamburg, Luruper Chaussee 149, 22761 Hamburg, Germany}
\author{G.C. Katsimiga}
\affiliation{Center for Optical Quantum Technologies, Department of Physics, University of Hamburg, 
Luruper Chaussee 149, 22761 Hamburg Germany}  
\author{Th. Busch}
\affiliation{OIST Graduate University, Onna, Okinawa 904-0495, Japan} 
\author{P. Schmelcher}
\affiliation{Center for Optical Quantum Technologies, Department of Physics, University of Hamburg, 
Luruper Chaussee 149, 22761 Hamburg Germany} \affiliation{The Hamburg Centre for Ultrafast Imaging,
Universit\"{a}t Hamburg, Luruper Chaussee 149, 22761 Hamburg, Germany}

\date{\today}

\begin{abstract} 

We study the ground state properties and nonequilibrium dynamics of two
spinor bosonic impurities immersed in a one-dimensional bosonic gas upon
applying an interspecies interaction quench.  For the ground state of two
non-interacting impurities we reveal signatures of attractive induced
interactions in both cases of attractive or repulsive interspecies
interactions, while a weak impurity-impurity repulsion forces the
impurities to stay apart.  Turning to the quench dynamics we inspect the
time-evolution of the contrast unveiling the existence, dynamical
deformation and the orthogonality catastrophe of Bose polarons.  We find
that for an increasing postquench repulsion the impurities reside in a
superposition of two distinct two-body configurations while at strong
repulsions their corresponding two-body correlation patterns show a
spatially delocalized behavior evincing the involvement of higher excited
states.  For attractive interspecies couplings, the impurities exhibit a
tendency to localize at the origin and remarkably for strong attractions
they experience a mutual attraction on the two-body level that is imprinted
as a density hump on the bosonic bath.

\end{abstract}

\maketitle

\section{Introduction}

Mobile impurities immersed in a quantum many-body (MB) environment become dressed by the excitations of the latter. This gives rise to the 
concept of quasiparticles, e.g. the polarons \cite{Massignan,Schmidt_rev}, which were originally introduced by Landau \cite{Landau,Pekar,Pekar1}. 
This dressing mechanism can strongly modify the elementary properties of the impurity atoms and lead to concepts such as effective mass and energy \cite{Grusdt_1D,Ardila_MC}, 
induced interactions \cite{induced_int_artem,Mistakidis_Fermi_pol} and attractively bound bipolaron states \cite{Bipolaron,Massignan,Schmidt_rev,Kevin}. 
Polaron states have been recently realized in ultracold atom experiments \cite{Scazza,Kohstall,Schirotzek}, which exhibit an unprecedented degree of controllability and, in particular,
allow to adjust the interaction between the impurities and the medium with the aid of Feshbach resonances \cite{Chin,Kohler}.
The spectrum of the quasiparticle excitations can be characterized in terms of radiofrequency and Ramsey spectroscopy \cite{Koschorreck,Kohstall,Cetina,Cetina_interferometry} and the 
trajectories of the impurities can be monitored via in-situ measurements \cite{Catani1,Fukuhara}. 
Experimentally Bose \cite{Jorgensen,Hu,Catani1,Fukuhara,Yan_bose_polarons} 
and Fermi \cite{Scazza,Koschorreck,Kohstall} polarons have been observed and these experiments confirmed the importance of higher-order correlations for the description of the polaronic 
properties. 
The experiments in turn have spurred additional several theoretical investigations which have aimed at describing different polaronic aspects \cite{Grusdt_approaches,Rath_approaches} by operating 
e.g.~within the Fr\"ohlich model \cite{Bruderer,Privitera,Casteels1,Casteels2,Kain}, effective Hamiltonian 
approximations \cite{Effect_hamilt,Effect_hamilt1,induced_int_artem,dynamics_Artem}, 
variational approaches \cite{Mistakidis_eff_mass,Mistakidis_orth_cat,Mistakidis_Fermi_pol,Jorgensen,Ardila_MC,Ardila_res_int}, 
renormalization group methods \cite{Grusdt_RG,Grusdt_strong_coupl,Grusdt_approaches} and the path integral formalism \cite{Tempere_path_int,Tempere_path2}. 

The focus of the majority of the above-mentioned theoretical studies have been the stationary properties of the emergent quasiparticle states 
for single impurities in homogeneous systems. 
However, the nonequilibrium dynamics of impurities is far less explored and is  expected to be dominated by correlation effects which build up in the 
course of the evolution \cite{dynamics_Artem,Mistakidis_orth_cat,Mistakidis_two_imp_ferm,Mistakidis_eff_mass,Grusdt_RG,Shchadilova,Kamar,Boyanovsky}. 
Existing examples include the observation of self-trapping phenomena \cite{Cucchietti_self_trap,Schecter_self_trap}, formation of dark-bright solitons \cite{Grusdt_1D,Mistakidis_two_imp_ferm}, 
impurity transport in optical lattices \cite{Cai_transp,Johnson_transp,Siegl,Theel}, orthogonality catastrophe events \cite{Mistakidis_orth_cat,catstrophe}, injection of 
a moving impurity into a gas of Tonks-Girardeau 
bosons \cite{Rutherford,Burovski_col,Lychkovskiy_col1,Lychkovskiy_col2,Meinert,Flutter,Flutter1,Gamayun_col} and the relaxation dynamics 
of impurities \cite{Lausch_col,Lausch_col1,Boyanovsky}. 
Besides these investigations, which have enabled a basic description of the quasiparticle states in different interaction regimes, a number of important questions remain open 
and a full theoretical understanding of the dynamics specifically of Bose polarons is still far from complete. 

A system of particular interest consists of two impurity atoms immersed in a Bose-Einstein condensate (BEC), where the underlying interactions 
between the impurities come into play. 
In such a system impurity-impurity correlations \cite{Huber_induced_cor,Bipolaron,Mistakidis_induce_int} can be induced by the BEC, even in the case where no direct interaction between the impurities is present. 
However, the competition between direct and induced interactions can also be expected to lead to interesting effects. 
It is therefore natural to investigate the dynamical response of the impurities with varying interspecies interactions (attractive or repulsive)  
and to identify in which regimes robustly propagating Bose polaron states exist \cite{Grusdt_approaches,Grusdt_RG,Shchadilova}. 
In addition it is interesting to study the existence of bound states between the impurities \cite{Bipolaron,Massignan}, the effect of strong correlation between the impurities on the orthogonality 
catastrophe  \cite{Mistakidis_orth_cat,catstrophe}, 
phase separation between the two atomic species \cite{mistakidis_phase_sep,Erdmann_phase_sep,Ao_phase_sep} and energy exchange processes \cite{Nielsen,Lampo}. 
Comparing the effects in systems with single and multiple impurities is an interesting task, as well as their theoretical interpretation in terms 
of the spin polarization (alias the contrast) which has not yet been analyzed in the case of two impurities and involves more energy channels compared to the case of a single impurity. 
For these reasons, we study in this work an interspecies interaction quench for two bosonic impurities overlapping with a harmonically trapped BEC. 
To address the correlated quantum dynamics of the bosonic multicomponent system we use the Multi-Layer Multi-Configuration Time-Dependent Hartree 
method for atomic mixtures (ML-MCTDHX) \cite{MLX,MLB1,MLB2}, which is a non-perturbative variational method that enables us to comprehensively capture interparticle correlations. 

In this work we start by studying the ground state of two non-interacting impurities in a bosonic gas and show that for an increasing 
attraction or repulsion they feature attractive induced interactions, a result that persists also for small bath sizes and heavy impurities \cite{induced_int_artem}. 
However, two weakly repulsively interacting impurities can experience a net repulsion for repulsive interspecies interactions. 

When quenching the multicomponent system, we monitor the time-evolution of the contrast and its spectrum \cite{Grusdt_approaches,Cetina_interferometry} 
for varying postquench interactions. 
We show that the polaron excitation spectrum depends strongly on the postquench interspecies interaction strength and the number of impurities 
while it is almost insensitive to the direct impurity-impurity interaction for the weak couplings considered herein. 
Additionally, a breathing motion of the impurities can be excited \cite{Sartori,Hannes} for weak postquench interspecies repulsions, while for stronger ones
a splitting of their single-particle density occurs. 
In this latter case a strong attenuation of the impurities motion results in the 
accumulation of their density at the edges of the bosonic gas and they mainly reside in a superposition of two distinct two-body configurations: 
the impurities either bunch on the same or on separate sides of the BEC, while the bath exhibits an overall
breathing motion. 
For attractive interspecies couplings, the impurities exhibit a breathing motion characterized by a beating pattern. 
The latter stems from the values of the impuritie's center-of-mass and relative coordinate breathing modes, whose frequency difference 
originates from the presence of attractive induced interactions. 
Additionally, the impurities possess a tendency to localize at the trap center, 
a behavior that becomes more pronounced for stronger attractions \cite{Mistakidis_inject_imp}. 
Strikingly, for strong attractive interspecies interactions we show that during the dynamics the impurities experience a mutual attraction on the two-body level and
the density of the bosonic bath develops a small amplitude hump at the trap center. 
We find that a similar dynamical response also takes place for two weakly repulsively interacting impurities but the involved time-scales are different. 
To interpret the observed dynamics of the impurities we invoke an effective potential picture that applies for weak 
couplings \cite{Mistakidis_orth_cat,Mistakidis_inject_imp,Hannes,Mistakidis_eff_mass}. 

Our work is structured as follows. 
Section \ref{sec:theory} presents our setup and introduces the correlation measures that are used to monitor the dynamics. 
In Sec. \ref{sec:ground} we address the ground state properties of the impurities for a wide range of interspecies interaction strengths. 
The emergent nonequilibrium dynamics triggered by an interspecies interaction quench is analyzed in detail in Sec. \ref{sec:quench}. 
In particular, we present the time-evolution of the contrast and the system's spectrum [Sec. \ref{sec:structure_interpretation}-\ref{sec:spectrum}] and 
study the full dynamics of the single-particle and two-body reduced density matrices for repulsive [Sec. \ref{sec:repulsive}] 
and attractive [Sec. \ref{sec:attractive}] postquench interactions. 
We summarize and discuss future perspectives in Section \ref{sec:conclusions}. 
Finally, Appendix \ref{sec:convergence} details our numerical simulation method and demonstrates the convergence properties.

\section{Theoretical Framework}\label{sec:theory}

\subsection{Hamiltonian and quench protocol}\label{sec:hamiltonian}

We consider a highly particle number imbalanced Bose-Bose mixture composed of $N_I=2$ bosonic impurities (I) possessing an additional pseudospin-$1/2$ degree 
of freedom \cite{Kasamatsu}, which are immersed in a bosonic gas of $N_B=100$ structureless bosons (B). 
Moreover, the mixture is assumed to be mass-balanced, namely $m_B=m_I\equiv m$ and each species is confined in the same one-dimensional external harmonic 
oscillator potential of frequency $\omega_B=\omega_I=\omega$. 
Such a system can be experimentally realized by considering e.g.~a $^{87}$Rb BEC where the majority species resides in the hyperfine state $\ket{F=2,m_F=1}$ 
and the pseudospin degree of freedom of the impurities refers for instance to 
the internal states $\ket{\uparrow}\equiv\ket{F=1,m_F=1}$ and $\ket{\downarrow}\equiv\ket{F=1,m_F=-1}$ \cite{Egorov,Alvarez}. 
Alternatively, it can be realized to a good approximation by a mixture of isotopes of $^{87}$Rb for the bosonic gas and two hyperfine 
states of $^{85}$Rb for the impurities. 
The underlying MB Hamiltonian of this system reads 
\begin{eqnarray}
\begin{split}
\hat{H} = \hat{H}^{0}_{B}+\sum_{a=\uparrow, \downarrow} \hat{H}^{0}_a+\sum_{a=\uparrow,\downarrow} \hat{H}_{aa}^{int}+\hat{H}_{\uparrow \downarrow}^{int}\\+\hat{H}_{BB}^{int}+\hat{H}_{BI}^{int}. 
\label{Htot_system}
\end{split}
\end{eqnarray}
The non-interacting Hamiltonian of the bosonic gas is $\hat{H}^{0}_{B}=\int dx~\hat{\Psi}^{\dagger}_{B} (x) \left( -\frac{\hbar^2}{2 m} \frac{d^2}{dx^2}  
+\frac{1}{2} m \omega^2 x^2 \right) \hat{\Psi}_{B}(x)$, while for the impurities it reads  
$\hat{H}^{0}_a=\int dx~\hat{\Psi}^{\dagger}_a(x) \left( -\frac{\hbar^2}{2 m} \frac{d^2}{dx^2}  
+\frac{1}{2} m \omega^2 x^2 \right) \hat{\Psi}_a(x)$ with $a=\left\{ \uparrow, \downarrow \right \}$ being the indices of the spin components. 
Here $\hat{\Psi}_{\sigma} (x)$ refers to the bosonic field-operator of either the bosonic gas ($\sigma=B$) or the impurity ($\sigma=a=\left\{ \uparrow, \downarrow \right \}$) atoms. 
Furthermore, we operate in the ultracold regime where $s$-wave scattering is the dominant interaction process. 
Therefore both the intra- and the intercomponent interactions can be adequately modeled by contact ones. 
The contact intraspecies interaction of the BEC component is modeled by 
$\hat{H}_{BB}^{int}=g_{BB} \int dx~\hat{\Psi}^{\dagger}_{B}(x) \hat{\Psi}^{\dagger}_{B}(x) \hat{\Psi}_{B} (x)\hat{\Psi}_{B}
(x)$ and between the impurities via $\hat{H}_{aa'}^{int}=g_{aa'}\int dx \hat{\Psi}_a^{\dagger}(x)\hat{\Psi}_{a'}^{\dagger}(x)\hat{\Psi}_{a'}(x)\hat{\Psi}_a(x)$ 
where either $a=a'=\uparrow,\downarrow$ or $a=\uparrow$, $a'=\downarrow$. 
Note also that we assume $g_{\uparrow \uparrow}=g_{\downarrow \downarrow}=g_{\uparrow \downarrow}\equiv g_{II}$. 
Most importantly, we consider that only the pseudospin-$\uparrow$ component of the impurities interacts with the bosonic gas while 
the pseudospin-$\downarrow$  is non-interacting. 
The resulting intercomponent interaction is $\hat{H}_{BI}^{int}=g_{BI}\int dx~\hat{\Psi}^{\dagger}_{B}(x) \hat{\Psi}^{\dagger}_{\uparrow}(x) \hat{\Psi}_{\uparrow} 
(x)\hat{\Psi}_{B}(x)$, where $g_{BI}\equiv g_{B\uparrow}$ and $g_{B\downarrow}=0$. 

In all of the above-mentioned cases, the effective one-dimensional coupling strength \cite{Olshanii} is given by  
${g_{\sigma \sigma'}} =\frac{{2{\hbar ^2}{a^s_{\sigma \sigma'}}}}{{\mu a_ \bot ^2}}{\left( {1 - {\left|{\zeta (1/2)} \right|{a^s_{\sigma \sigma'}}}
/{{\sqrt 2 {a_ \bot }}}} \right)^{ -1}}$, where $\sigma,\sigma'=B, \uparrow,\downarrow$ and $\mu=\frac{m}{2}$ is the reduced mass. 
The transversal length scale is ${a_\bot } = \sqrt{\hbar /{\mu{\omega _ \bot }}}$ with ${{\omega _ \bot }}$ being the transversal confinement frequency 
and ${a^s_{\sigma \sigma'}}$ denotes the three-dimensional $s$-wave scattering length within ($\sigma=\sigma'$) or between ($\sigma \neq \sigma'$) 
the components. 
In a corresponding experiment, $g_{\sigma\sigma'}$ can be tuned either via ${a^s_{\sigma \sigma'}}$ with the aid of Feshbach 
resonances \cite{Kohler,Chin} or by adjusting ${{\omega _ \bot }}$ using confinement-induced resonances \cite{Olshanii}. 
In the following, the MB Hamiltonian of Eq. (\ref{Htot_system}) is rescaled with respect to $\hbar  \omega$.  
As a consequence, length, time, and interaction strengths are given in units of $\sqrt{\frac{\hbar}{m \omega}}$, $\omega^{-1}$ and 
$\sqrt{\frac{\hbar^3 \omega}{m}}$ respectively. 

To study the quench dynamics, the above-described multicomponent system is initially prepared in its ground state configuration for fixed 
$g_{BB}=0.5$ and $g_{BI}=0$ and either $g_{II}=0$ or $g_{II}=0.2$. 
In this way, the case of two non-interacting and that of weakly interacting impurities are 
investigated. 
This initial (ground) state emulates a system prepared in the $\ket{1,-1}=\ket{\downarrow}_1\otimes\ket{\downarrow}_2$ configuration 
for the spin degree of freedom i.e. where the impurity-BEC interaction is zero. 
Note that the spinor part of the wavefunction is expressed in the basis of the total spin i.e.~$\ket{S,S_z}$ \cite{Tannoudji}. 
Accordingly, the spatial part $\ket{\Psi^0_{BI}}$ of the ground state of the system obeys the following eigenvalue equation 
$\left(\hat{H}-\hat{H}_{BI}\right)\ket{\Psi^0_{BI}} \ket{1,-1}=E_0\ket{\Psi^0_{BI}}\ket{1,-1}$, with $E_0$ being the corresponding eigenenergy and 
$\hat{H}_{BI}\ket{\Psi^0_{BI}} \ket{1,-1}=0$. 
To trigger the dynamics we carry out an interspecies interaction quench from $g_{BI}=0$ to a finite positive or negative value of $g_{BI}$ at $t=0$ and 
monitor the subsequent time-evolution. 
In a corresponding experiment, this quench protocol can be implemented by using a radiofrequency $\pi/2$ pulse with an exposure time much smaller than $\omega^{-1}$ \cite{Cetina_interferometry}. 
The pulse acts upon the spin degree of freedom of the impurity, which maps the pseudospin-$\downarrow$ impurities to the 
superposition state $\ket{\psi_S}_i\equiv \frac{\ket{\uparrow}_i+\ket{\downarrow}_i}{\sqrt{2}}$ with $i=1,2$ \cite{Cetina}. 
The corresponding MB wavefunction of the system, $\ket{\Psi(t)}=e^{-i\hat{H} t/\hbar} \big[\ket{\Psi^0_{BI}} (\ket{\psi_S}_1 \otimes \ket{\psi_S}_2)\big]$, is then given by 
\begin{equation}
\begin{split}
\ket{\Psi(t)}&=  \frac{1}{\sqrt{2}} e^{-i\hat{H} t/\hbar} 
\big[\ket{\Psi^0_{BI}} \ket{1,0}\big]+ \frac{1}{2} \big (e^{-iE_0 t/\hbar} \\& \times \ket{\Psi^0_{BI}} \ket{1,-1}+ e^{-i\hat{H}t/\hbar} \ket{\Psi_{BI}^0} \ket{1,1}\big).	
\label{Eq:2}
\end{split}
\end{equation}
The setup and processes addressed in our work can be experimentally realized utilizing radiofrequency spectroscopy \cite{Jorgensen,Hu,Cetina,Shchadilova,Mistakidis_Fermi_pol} 
and Ramsey interferometry \cite{Cetina}.

\subsection{Many-body wavefunction ansatz} \label{sec:many_body_ansatz}

To calculate the stationary properties and to track the MB nonequilibrium quantum dynamics of the 
multicomponent bosonic system discussed above we employ the ML-MCTDHX method \cite{MLX,MLB1,MLB2}. 
This is an ab-initio variational method for solving the time-dependent MB Schr{\"o}dinger equation of atomic 
mixtures and it is based on the expansion of the total MB wavefunction with respect to a time-dependent and variationally 
optimized basis tailored to capture both the intra- and the interspecies correlations of a multicomponent system
\cite{Mistakidis_orth_cat,mistakidis_phase_sep,Koutentakis_prob,Katsimiga_DB}.  

To include the interspecies correlations, the MB wavefunction ($|\Psi(t)\rangle$) is first expanded in terms of $D$ distinct species 
functions, $|\Psi^{\sigma}_i(t)\rangle$, for each component $\sigma=B,I$, and then 
expressed according to a truncated Schmidt decomposition \cite{Horodecki} of 
rank $D$, namely    
\begin{equation} 
    |\Psi(t)\rangle=\sum_{k=1}^D \sqrt{\lambda_k(t)} |\Psi^{\rm B}_k(t)\rangle|\Psi^{\rm I}_k(t)\rangle.  
    \label{eq:wfn}
\end{equation} 
Here the time-dependent expansion coefficients $\lambda_k(t)$ are the Schmidt weights and will be referred to in the following as the 
natural populations of the $k$-th species function. 
Evidently, the system is entangled \cite{Roncaglia} or interspecies correlated when at least two different $\lambda_k(t)$ 
possess a nonzero value. 
If this is not the case, i.e. for  $\lambda_1(t)=1$, $\lambda_{k>1}(t)=0$, the wavefunction is a direct product of two states. 

Therefore, in order to account for intraspecies correlations, each of the above-mentioned species functions is 
expressed as a linear superposition of time-dependent number-states, $|\vec{n} (t) \rangle^{\sigma}$, with time-dependent 
coefficients $A^{\sigma}_{i;\vec{n}}(t)$ as
\begin{equation}
    | \Psi_i^{\sigma} (t) \rangle =\sum_{\vec{n}} A^{\sigma}_{i;\vec{n}}(t) | \vec{n} (t) \rangle^{\sigma}.  
    \label{eq:number_states}
\end{equation} 
Each number state $|\vec{n} (t) \rangle^{\sigma}$ is a permanent building upon $d^{\sigma}$ time-dependent variationally 
optimized single-particle functions (SPFs) $\left|\phi_l^{\sigma} (t) \right\rangle$, $l=1,2,\dots,d^{\sigma}$ with occupation
numbers $\vec{n}=(n_1,\dots,n_{d^{\sigma}})$. 
Consecutively, the SPFs are expanded on a time-independent primitive basis. 
The latter refers to an $\mathcal{M}$ dimensional discrete variable representation (DVR) for the majority species and it is denoted by 
$\lbrace \left| k \right\rangle \rbrace$. 
For the impurities this corresponds to the tensor product $\lbrace \left| k,s \right\rangle \rbrace$ 
of the DVR basis for the spatial degrees of freedom and the two-dimensional pseudospin-$1/2$ basis $\{\ket{\uparrow}, \ket{\downarrow}\}$. 
Accordingly, each SPF of the impurities is a spinor wavefunction of the form  
\begin{equation}
    | \phi^{\rm I}_j (t) \rangle= \sum_{k=1}^{\mathcal{M}}\big( B^{{\rm
    I}}_{jk \uparrow}(t) \ket{k} \ket{\uparrow}+B^{{\rm I}}_{jk \downarrow}(t) \ket{k} \ket{\downarrow}\big), 
    \label{eq:spfs}
\end{equation}
with $B^{{\rm I}}_{j k \uparrow}(t)$ [$B^{{\rm I}}_{j k \downarrow}(t)$] being the time-dependent expansion coefficients 
of the pseudospin-$\uparrow$ [$\downarrow$] (see also Refs. \cite{Mistakidis_orth_cat,Koutentakis_prob} 
for a more detailed discussion).

The time-evolution of the ($N_B+N_I$)-body wavefunction $\left|\Psi(t) \right\rangle$ governed by the Hamiltonian of Eq.~(\ref{Htot_system})  
is obtained via solving the so-called ML-MCTDHX equations of motion \cite{MLX}. 
The latter are determined by utilizing e.g.~the Dirac-Frenkel \cite{Frenkel,Dirac} variational principle for the generalized ansatz introduced 
in Eqs.~(\ref{eq:wfn}), (\ref{eq:number_states}) and (\ref{eq:spfs}). 
This procedure results in a set of $D^2$ linear differential equations of motion for the $\lambda_k(t)$ coefficients which are 
coupled to $D(\frac{(N_B+d^B-1)!}{N_B!(d^B-1)!}+\frac{(N_I+d^I-1)!}{N_I!(d^I-1)!})$ 
nonlinear integrodifferential equations for the species functions and $d^B+d^I$ nonlinear integrodifferential equations for the SPFs. 

A main aspect of the ansatz outlined above is the expansion of the system's MB wavefunction
with respect to a time-dependent and variationally optimized basis.  The latter allows to
efficiently take into account the intra- and intercomponent correlations of the system using a
computationally feasible basis size. In the present case the Bose gas consists of a large 
number of weakly interacting particles and therefore its intracomponent correlations are
suppressed. 
As a consequence they can be adequately captured by employing a small number of orbitals, $d^B<4$. 
Additionally, the number of impurities, $N_I<3$, is small giving rise to a small number of integrodifferential equations
allowing us to employ many orbitals, $d_I$, and thus account for strong impurity-impurity
and impurity-BEC correlations.  Therefore, the number of the resulting equations of motion
that need to be solved is numerically tractable.  Since our method is variational, its validity is
determined upon examining its convergence.  For details on the precision of our simulations see
Appendix \ref{sec:convergence}.

\subsection{Correlation measures}\label{sec:observables}

To study the quench-induced dynamics of each species at the single-particle level we calculate 
the one-body reduced density matrix for each species \cite{Naraschewski,density_matrix} 
\begin{equation}
\rho_\sigma^{(1)}(x,x';t)=\langle\Psi(t)|\hat{\Psi}_{\sigma}^{\dagger}(x)\hat{\Psi}_{\sigma}(x')|\Psi(t)\rangle.\label{eq:single_particle_density_matrix}
\end{equation}
Here, $\hat{\Psi}_{\sigma}(x)$ is the $\sigma$-species bosonic field operator acting at position $x$ and satisfying 
the standard bosonic commutation relations \cite{Pethick_book}. 
For simplicity, we will use in the following the one-body densities for each species i.e.~$\rho_\sigma^{(1)}(x;t)\equiv\rho_\sigma^{(1)}(x,x'=x;t)$, 
which is a quantity that is experimentally accessible via averaging over a sample of single-shot images \cite{Bloch_review,Jochim_resolved,mistakidis_phase_sep}. 
We remark that the eigenfunctions and eigenvalues of $\rho_\sigma^{(1)}(x,x';t)$ are termed natural orbitals $\varphi^{\sigma}_i(x;t)$ 
and natural populations $n^{\sigma}_i(t)$ \cite{mistakidis_phase_sep,MLX} respectively. 
In this sense, each bosonic subsystem is called intraspecies correlated if more than a single natural population possess a non-zero contribution. 
Otherwise, i.e.~ for $n_1^{\sigma}(t)=1$ and $n_{i>1}^{\sigma}(t)=0$, the corresponding subsystem is said to be fully coherent and the MB wavefunction 
[Eqs. (\ref{eq:wfn}), (\ref{eq:spfs})] reduces to a mean-field product ansatz \cite{emergent,stringari}. 

To unveil the role of impurity-impurity correlations following the interspecies interaction quench we calculate the time-evolution of the 
corresponding diagonal of the two-body reduced density matrix 
\begin{equation}
\begin{split}
\rho^{(2)}_{aa'}(x_1,x_2;t)=\bra{\Psi(t)}\hat{\Psi}_{a}^{\dagger}(x_{1}) \hat{\Psi}_{a'}^{\dagger}(x_{2}) \hat{\Psi}_{a'}(x_{2}) \\ \times \hat{\Psi}_{a}(x_{1})\ket{\Psi(t)}, 
\label{eq:two_body_density}
\end{split}
\end{equation}
where $a,a'=\uparrow,\downarrow$. 
The two-body reduced density matrix refers to the probability of finding simultaneously one pseudospin-$a$ boson at $x_1$ and 
a pseudospin-$a'$ boson at $x_2$ \cite{mistakidis_phase_sep,Erdmann_phase_sep}. 
Moreover, it provides insights into the spatially resolved dynamics of the two impurities with respect to one another. 
Indeed, the impurities are dressed by the excitations of the bosonic gas forming quasiparticles which in turn can move independently 
or interact, and possibly form a bound state \cite{induced_int_artem,Bipolaron,Klein,Mistakidis_two_imp_ferm}. 

To capture the emerging effective interactions between the two bosonic impurities we monitor 
their relative distance \cite{Mistakidis_Fermi_pol,Mistakidis_two_imp_ferm} given by  
\begin{equation}
\label{eq:distance}
\begin{split}
\braket{r_{aa}(t)}=\frac{\int dx_1 dx_2 |x_1-x_2| \rho^{(2)}_{aa}(x_1,x_2;t)}{
\braket{\Psi(t)|\hat{N}_{a} \left(\hat{N}_{a} -1\right)|\Psi(t)}}.
\end{split}
\end{equation} 
Here, $\hat{N}_{a}$ with $a=\uparrow,\downarrow$ is the number operator that measures the number of bosons in the spin-$a$ state. 
Experimentally, $\braket{r_{aa}(t)}$ can be probed via {\it in-situ} spin-resolved single-shot measurements on the spin-$a$ state \cite{Jochim_resolved}. 
More precisely, each image gives an estimate of $\braket{r_{aa}(t)}$ between the bosonic impurities if their position uncertainty is assured to 
be adequately small \cite{Jochim_resolved}. 
Subsequently, $\braket{r_{aa}(t)}$ is obtained by averaging over several such images.

\section{Induced interactions in the ground state of two bosonic impurities}\label{sec:ground}

Before investigating the nonequilibrium dynamics of the two bosonic impurities immersed in a BEC it is instructive to first analyze the 
ground state of two impurities interacting with the bosonic medium for varying interspecies interactions $g_{B\uparrow}$ ranging from attractive to repulsive. 
Note that such a configuration corresponds in our case to two impurities residing in the pseudospin-$\uparrow$ state since only this state is interacting with the bath 
[see also Eq. (\ref{Htot_system})]. 
The aim of this study is to reveal the presence of induced impurity-impurity interactions mediated by the bath. 
As discussed in Section \ref{sec:hamiltonian}, the mass-balanced multicomponent bosonic system consists of two impurities $N_I=2$ immersed in 
a MB bath of $N_B=100$ atoms with $g_{BB}=0.5$ and it is externally confined in a harmonic oscillator 
potential of frequency $\omega=1$. 
Later on, also the mass-imbalanced and the few-body ($N_B=10$) scenaria will be investigated. 
Below we consider either two non-interacting ($g_{II}=0$) 
or two weakly interacting impurities ($g_{II}=0.2$). 
To obtain the interacting ground state of the system as described by the Hamiltonian of Eq. (\ref{Htot_system}) we employ either imaginary time propagation 
or improved relaxation \cite{MLX,MLB1} within ML-MCTDHX. 
\begin{figure*}[ht]
  	\includegraphics[width=0.78\textwidth]{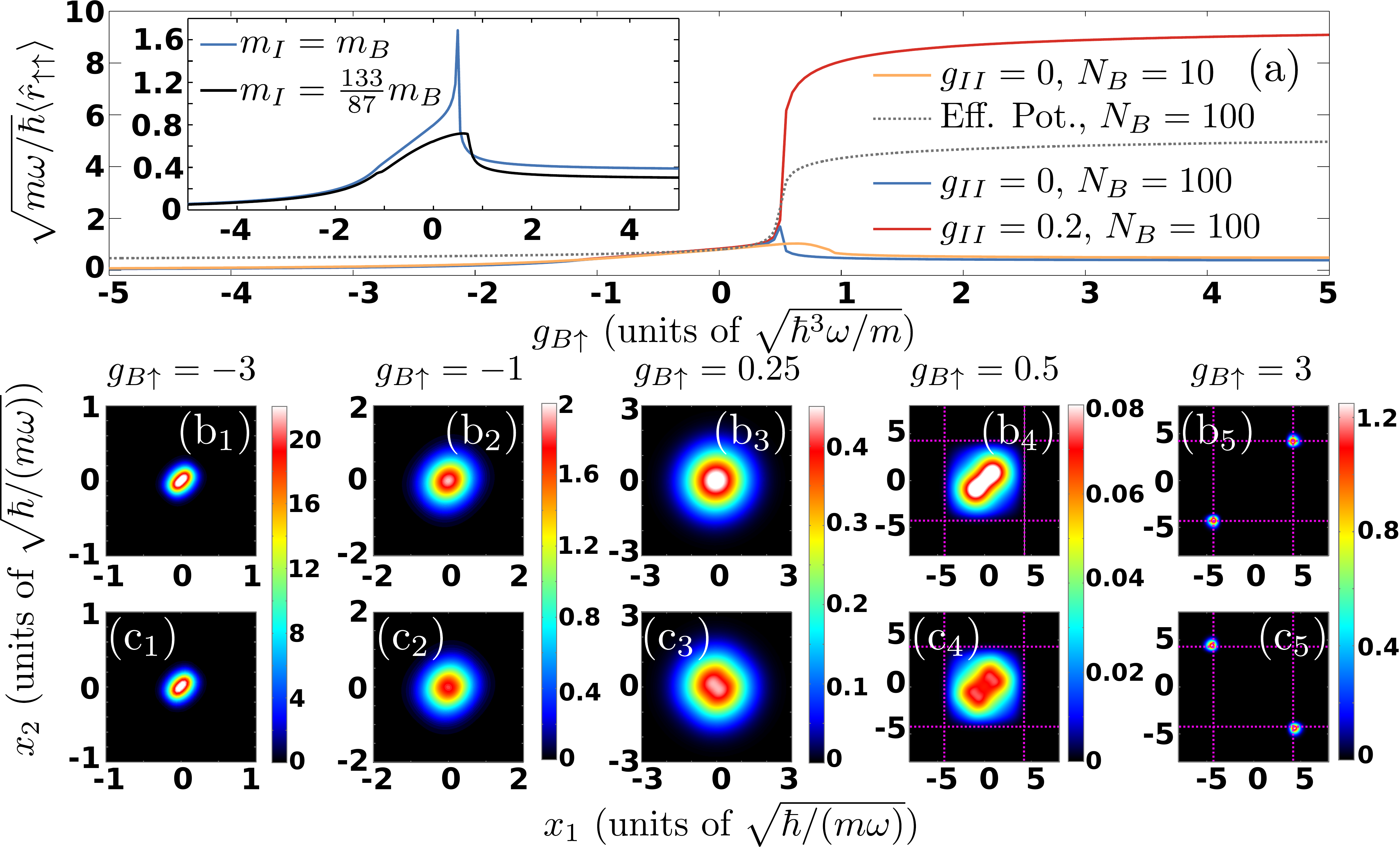}
  	\caption{(a) Relative distance, $\braket{r_{\uparrow \uparrow}}$, between the two bosonic 
  	impurities residing in the pseudospin-$\uparrow$ state for varying bath pseudospin-$\uparrow$ interaction 
  	strength. 
  	The cases of two non-interacting ($g_{II}=0$), weakly interacting ($g_{II}=0.2$) impurities as 
  	well as few- and many bath particles are shown (see legend) for a mass-balanced system $m_I=m_B$. 
  	$\braket{r_{\uparrow \uparrow}}$ from the effective potential picture of Eq. (\ref{eq:eff_pot_stat}) for two non-interacting bosonic 
  	impurities is also illustrated (see legend) with respect to $g_{B\uparrow}$. 
  	Inset illustrates $\braket{r_{\uparrow \uparrow}}$ of two non-interacting impurities in the case of a mass-balanced ($m_I=m_B$) 
  	and a mass-imbalanced ($m_I\approx1.53m_B$) system with respect to $g_{B\uparrow}$.
  	The corresponding two-body reduced matrix of the ground state of the two pseudospin-$\uparrow$ 
  	($b_1$)-($b_5$) non-interacting and ($c_1$)-($c_5$) interacting ($g_{II}=0.2$) impurities for 
  	different interspecies interactions (see legends). 
  	In ($b_1$)-($b_5$) and ($c_1$)-($c_5$) the mixture consists of $N_B=100$ bosons and $N_I=2$ 
  	bosonic impurities.  
  	Also, in ($b_4$), ($b_5$), ($c_4$) and ($c_5$) the dashed magenta lines indicate the location 
  	of the Thomas-Fermi radius of the bosonic gas. 
  	In all cases $g_{BB}=0.5$ and the system is trapped in a harmonic oscillator potential with $\omega=1$.}
  	\label{fig:ground} 
\end{figure*} 

The relative distance [Eq. (\ref{eq:distance})] between the two impurities as well as their two-body reduced density matrix [Eq.~(\ref{eq:two_body_density})] 
for different values of $g_{B\uparrow}$ are shown in Fig. \ref{fig:ground}. 
Focusing on the case of two non-interacting impurities, $g_{II}=0$, we see that for larger attractions the relative distance between the impurities 
decreases (see Fig, \ref{fig:ground} (a)) and converges towards a constant value i.e.~$\braket{r_{\uparrow \uparrow}}\approx 0.1$ for $g_{B\uparrow}<-2$. 
The decrease in  $\braket{r_{\uparrow \uparrow}}$ for $-2<g_{B\uparrow}<0$ implies that the impurities effectively experience an attraction with respect to one another.
This attraction is a manifestation of the attractive induced interactions mediated by the bosonic gas since $g_{II}=0$ \cite{induced_int_artem}. 
The impurities reside together in the vicinity of the trap center since $\rho^{(2)}_{\uparrow \uparrow}(-1<x_1<1,-1<x_2<1)$ is predominantly populated [see Fig.~\ref{fig:ground} ($b_2$)]. 
Additionally, for $g_{B\uparrow}<-2$, where $\braket{r_{\uparrow \uparrow}}$ become approximately constant, the impurities come very close with respect to one another. 
Here, the corresponding $\rho^{(2)}_{\uparrow \uparrow}(x_1,x_2)$ shrinks along its anti-diagonal and its diagonal becomes elongated [see Fig.~\ref{fig:ground} ($b_1$)], 
which is indicative of a bound state having formed between the impurities known as a bipolaron state \cite{induced_int_artem,Bipolaron,Klein}. 

Turning to weak interspecies repulsions $0<g_{B\uparrow}<0.5$ we find that 
$\braket{r_{\uparrow \uparrow}}$ slightly increases [see Fig.~\ref{fig:ground} (a)] 
while the two impurities reside close to the trap center [see Fig.~\ref{fig:ground} ($b_3$)]. 
It is important to mention that this increase in $\braket{r_{\uparrow \uparrow}}$ does 
not directly imply that the impurities experience a weak repulsion mediated by the bosonic bath. 
Indeed, by neglecting all correlations between the impurities, 
i.e.~by substituting $\rho^{(2)}_{\uparrow \uparrow}(x_1,x_2)=\rho^{(1)}_{\uparrow}(x_1)\rho^{(1)}_{\uparrow}(x_2)/2$ into 
$\braket{r_{\uparrow \uparrow}}$ we find the same tendency of $\braket{r_{\uparrow \uparrow}}$ with even slightly larger values (see also the discussion below). 
Since in the limit of the non-correlated case there are no induced interactions, 
the fact that $\braket{r_{\uparrow \uparrow}}$ is smaller when correlations are taken into account 
means that the impurities still feel an effective attractive force. 
Note that for the other interaction regimes presented herein such an unexpected behavior of $\braket{r_{\uparrow \uparrow}}$ does not occur as it can 
also be deduced by the corresponding two-body spatial configurations building upon $\rho^{(2)}_{\uparrow \uparrow}(x_1,x_2)$ (see below). 
Furthermore, it can be seen that at $g_{B\uparrow}=g_{BB}=0.5$, where the miscibility/immiscibility transition between the impurity and the BEC takes place \cite{Ao_phase_sep,mistakidis_phase_sep}, 
the behavior of $\braket{r_{\uparrow \uparrow}}$ is suddenly altered. 
Indeed for $g_{B\uparrow}\geq0.5$, $\braket{r_{\uparrow \uparrow}}$ shows a decreasing tendency which indicates the presence of attractive induced interactions 
between the impurities.  
In particular, for $0.5\leq g_{B\uparrow}<1.1$, $\braket{r_{\uparrow \uparrow}}$ reduces and the impurities tend to bunch together at the same location. 
This can be confirmed by the fact that $\rho^{(2)}_{\uparrow \uparrow}(x_1,x_2)$ shows a populated elongated diagonal as depicted 
in Fig.~\ref{fig:ground} ($b_4$) for $g_{B\uparrow}=0.5$. 
Moreover for stronger repulsions $g_{B\uparrow}>1.1$, $\braket{r_{\uparrow \uparrow}}$ remains almost constant. 
Especially so for $g_{B\uparrow}>1.5$, where the two impurities residing either on the left or the right edge of the Thomas-Fermi profile of the BEC. 
The latter can be evidenced in Fig.~\ref{fig:ground} ($b_5$) by the two strongly populated spots appearing at $x_1\approx x_2\approx \pm R_{TF}$ 
with $R_{TF}$ denoting the Thomas-Fermi radius. 

In view of the results of Ref. \cite{Mistakidis_orth_cat} it is tempting to interpret our above findings in terms of an effective 
potential, $V_{eff}(x;g_{BI})$. 
A valid candidate for such a potential can be constructed as
\begin{equation}
 V_{eff}(x;g_{BI})=\frac{1}{2} m_I \omega^2 x^2 + g_{BI} \rho^{(1)}_B(x;g_{BI}=0),
 \label{eq:eff_pot_stat}
\end{equation}
where $\rho^{(1)}_B(x;g_{BI}=0)$ refers to the equilibrium density of the BEC for $g_{BI}=0$. Equation (\ref{eq:eff_pot_stat}) implies
that $\rho^{(1)}_B(x;g_{BI}=0)$ acts on the impurities just as an additional repulsive ($g_{BI}>0$) or attractive ($g_{BI}<0$) potential
on top of the externally imposed parabolic trap. 
It is noteworthy that the simplification of the impurity problem provided by Eq. (\ref{eq:eff_pot_stat})
neglects several phenomena that might be important for the description of the ground state of the impurity system. First, the renormalization
of the impurity's mass, $m_I \to m_I^{eff}$ by the coupling with its environment is neglected and, 
most importantly, the possible emergence of
induced interactions is not contained in Eq.~(\ref{eq:eff_pot_stat}), due to the absence of two-body terms. 
The latter are extremely important for the description of $\rho^{(2)}_{\uparrow \uparrow}(x_1,x_2)$. 
Indeed, within $V_{eff}(x;g_{BI})$ no deformations can appear in the antidiagonal of the two-body density of the impurities which dictates their relative distance. 
This result is in contrast to the one obtained within the full MB Hamiltonian [Eq. (\ref{Htot_system})] shown in Figs. 1(b$_1$)-1(b$_5$). 

To provide an estimate of the quantitative error obtained by the approximation of Eq. (\ref{eq:eff_pot_stat}) we include in Fig. 1(a), also the
results for $\langle r_{\uparrow \uparrow} \rangle$ within the effective potential picture. It is evident that when using $V_{eff}(x;g_{BI})$,  $\langle
r_{\uparrow \uparrow} \rangle$ is always larger than the corresponding full MB result for $g_{BI} \neq 0$ . This effect is particularly
pronounced for $g_{BI} > 0.5$ where $\langle r_{\uparrow \uparrow} \rangle$ within Eq. (\ref{eq:eff_pot_stat}) exhibits an increasing
tendency instead of a decreasing one with $g_{BI}$. Such an effect can be attributed to the vanishing off-diagonal elements of $\rho^{(2)}_{\uparrow \uparrow}(x_1,x_2)$ which cannot be captured within 
$V_{eff}(x;g_{BI})$, as in the latter case $\rho^{(2)}_{\uparrow \uparrow}(x_1,x_2)=\rho^{(1)}_{\uparrow}(x_1) \rho^{(1)}_{\uparrow}(x_2)/2$. 
Indeed, the large impurity-impurity interactions within this regime render the effective potential incapable of describing the ground state of the bath impurity system within this interaction regime.
Similarly, for $g_{BI}<-2$, $\langle r_{\uparrow \uparrow} \rangle$ using the effective potential is significantly larger than the corresponding MB result, which can 
be attributed to the prominent role of induced interactions 
in the formation of the bipolaron state \cite{Bipolaron}. 

Considering a smaller bath consisting of $N_B=10$ atoms does not significantly alter the ground state properties of the two non-interacting bosonic impurities. 
Here, $\braket{r_{\uparrow \uparrow}}$ [Fig. \ref{fig:ground} (a)] exhibits a similar behavior as for $N_B=100$ atoms, with  
the most notable difference occurring in the region of $g_{B\uparrow}\approx g_{BB}$ where a smoother decrease occurs when compared to the $N_B=100$ case. 
The value for which the distance becomes constant is also shifted to larger values when $N_B=10$. 
These differences can be qualitatively understood within a corresponding effective potential picture which we will discuss 
in Section \ref{sec:density_repulsive}, see Eq.~(\ref{effective_potential_repulsive}) and the remark \cite{comment_few}.  
\begin{figure*}[ht]
  	\includegraphics[width=0.78\textwidth]{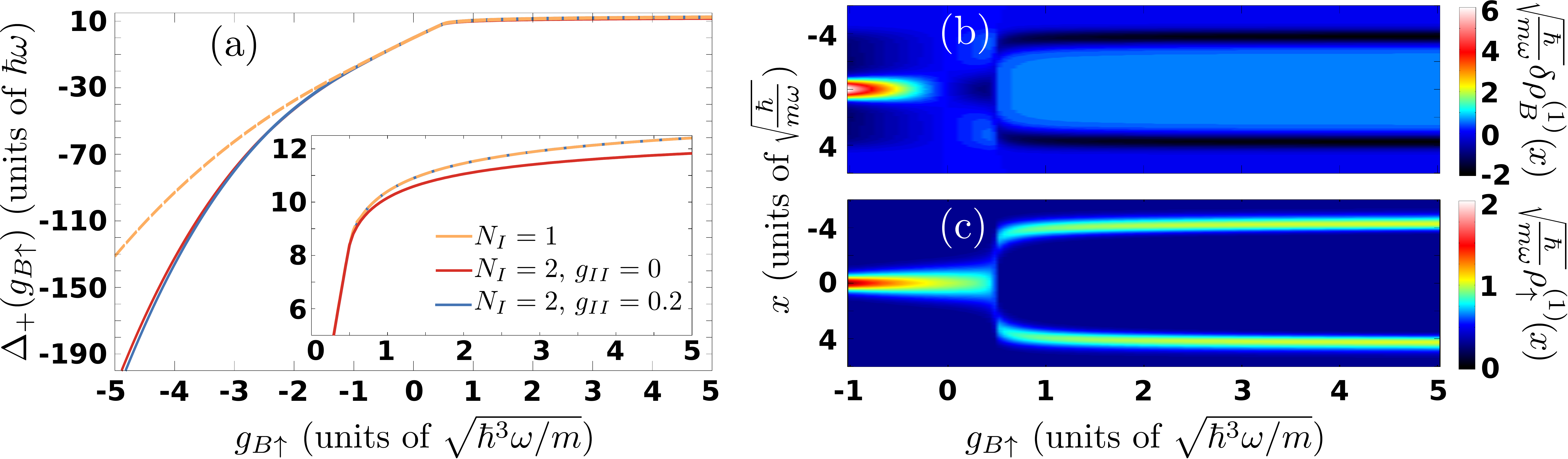}
        \caption{(a) Position of the polaronic resonances, 
                $\Delta_+^{N_I}(g_{B\uparrow})$, with varying $g_{B\uparrow}$ for $N_I=1$ and $N_I=2$ bosonic
                non-interacting and weakly interacting impurities (see legend). 
                Inset: $2<\Delta_+^{N_I}(g_{B\uparrow})<12.5$ for $g_{B\uparrow}>0$.
                (b) Deformation of the BEC ground state density measured via $\delta
                \rho^{(1)}_B(x;g_{B\uparrow})=\rho^{(1)}_B(x;g_{B\uparrow})-\rho^{(1)}_B(x;0)$ with
                respect to $g_{B\uparrow}$ for $N_I=2$ and $g_{II}=0$.  (c) Ground state
                one-body density of two non-interacting impurities as a function of $g_{B\uparrow}$.
                In all cases the bath consists of $N_B=100$ bosons with $g_{BB}=0.5$.}
  	\label{fig:energies} 
\end{figure*} 

A similar to the above-described overall phenomenology of the two non-interacting bosonic impurities for a varying $g_{B\uparrow}$ is also observed for the case of 
heavier impurities as can be seen in the inset of Fig.~\ref{fig:ground} (a). 
Here we consider a $^{87}$Rb bosonic gas and two $^{133}$Cs impurities prepared e.g.~in the hyperfine states $\Ket{F=1, m_F=0}$ and $\Ket{F=3, m_F=2}$ respectively and being 
both confined in the same external harmonic oscillator \cite{Hohmann_Rb_Cs,Spethmann_Rb_Cs}. 
Compared to the mass-balanced scenario the behavior of $\braket{r_{\uparrow \uparrow}}$ around $g_{B\uparrow}\approx g_{BB}$  
becomes somewhat smoother and the maximum value is also slightly shifted to larger interaction strengths. 
Another conclusion that can be drawn, is that heavier impurities prefer to remain closer to each other compared to the lighter ones, since 
$\braket{r_{\uparrow \uparrow}}$ has smaller values in the former than in the latter case. 
As a consequence we can infer that heavy impurities experience stronger attractive induced interactions than light ones. 
These differences can also be explained in terms of the effective potential picture which will be introduced 
in section \ref{sec:density_repulsive}, see also remark \cite{comment_imbalance}. 

When a weak intraspecies repulsion among the impurities is introduced, $g_{II}=0.2$, see Fig.~\ref{fig:ground}(a), 
the ground state properties remain the same for attractive $g_{B\uparrow}$ but change fundamentally in the repulsive regime. 
Indeed $\braket{r_{\uparrow \uparrow}}$ decreases for an increasing interspecies attraction, signifying an induced attraction between the impurities 
despite their repulsive mutual interaction, until it becomes constant for $g_{B\uparrow}<-2$. 
More specifically, for $-2<g_{B\uparrow}<0$ the impurities are likely to remain close to the trap center [see Fig.~\ref{fig:ground}($c_2$)] 
where $\rho^{(2)}_{\uparrow \uparrow}(-1<x_1<1,-1<x_2<1)$ is predominantly populated. 
Furthermore, for $g_{B\uparrow}<-2$ the impurities bunch together at a fixed distance [Fig.~\ref{fig:ground}(a)] and 
the two-body reduced density matrix becomes elongated along its diagonal [see Fig. \ref{fig:ground}($c_1$)], 
suggesting the formation of a bound state similar to the $g_{II}=0$ case. 
However, for $g_{B\uparrow}>0$, $\braket{r_{\uparrow \uparrow}}$ exhibits an overall increasing tendency, which indicates that the two impurities are located 
mainly symmetrically around the trap center. 
This latter behavior can be directly deduced by the relatively wide distribution of the
anti-diagonal of their two-body reduced density 
matrix [see Figs.~\ref{fig:ground} ($c_3$) and ($c_4$) for $0<g_{B\uparrow}<1$]. 
Moreover, and in sharp contrast to the $g_{II}=0$ case, for $g_{B\uparrow}>1$ the impurities acquire a large fixed distance and 
in particular can be found to reside one at the left and the other at the right edge of the BEC. 
This configuration of the impurities can be seen from the fact that solely off-diagonal elements of $\rho^{(2)}_{\uparrow \uparrow}(x_1,x_2)$ exist 
in Fig.~\ref{fig:ground}($c_5$) for $g_{B\uparrow}=3$. 
Finally, it is worth mentioning that for two weakly repulsive impurities the induced effective attraction can never overcome their direct s-wave interaction 
for $g_{B\uparrow}>0$. 

To further support the existence of attractive induced interactions between the two
impurities we study the ground state energy of the system for varying $g_{B\uparrow}$.  In
particular, we calculate the expected position of the polaronic resonances
\cite{Mistakidis_Fermi_pol} namely
$\Delta_+^{N_I}(g_{B\uparrow})=[E(N_I,g_{B\uparrow})-E(N_I,g_{B\uparrow}=0)]/N_I$, where
$E(N_I,g_{B\uparrow})$ is the energy of the system for $N_I$ impurities at interaction
$g_{B\uparrow}$ [Fig. \ref{fig:energies} (a)]. 
As it can be seen, for both, $N_I=1$ and $N_I=2$, the resonance position $\Delta_+^{N_I}(g_{B\uparrow})$ 
increases for a larger $g_{B\uparrow}$ and it takes negative and
positive values for attractive and repulsive interactions, respectively. 
Moreover, in the $N_I=2$ scenario $\Delta_+^{N_I}(g_{B\uparrow})$ is found to be negatively shifted when
compared to the corresponding $N_I=1$ case for $g_{II} \neq 0$. 
This behavior indicates the presence of attractive induced interactions for both attractive and repulsive Bose polarons
\cite{Huber_induced_cor,Bipolaron,induced_int_artem}.  Focusing on $g_{II}=0.2$ and
$g_{B\uparrow}<0$ a small decrease of $\Delta_+^{N_I}(g_{B\uparrow})$ occurs when compared to
the $g_{II}=0$ case showing that attractive induced interactions become more pronounced when
direct s-wave impurity-impurity repulsions are involved.  However, for repulsive polarons i.e.
$g_{B\uparrow}>0$ the presence of s-wave impurity-impurity interactions counteracts the effect of
attractive induced interactions and accordingly $\Delta_+^{N_I}(g_{B\uparrow})$ is almost the same
for $N_I=2$, $g_{II}=0.2$ and $N_I=1$, see the inset of Fig. \ref{fig:energies} (a). 

The underlying mechanism behind the above-mentioned impurity-impurity induced interactions can be qualitatively understood as follows. 
For attractive $g_{B\uparrow}$ the presence of impurities gives rise to a small density enhancement of the BEC in the vicinity of their spatial position. 
This effect is captured by the deformation of the BEC density quantified by $\delta \rho^{(1)}_B(x;g_{B\uparrow})=\rho^{(1)}_B(x;g_{B\uparrow})-\rho^{(1)}_B(x;0)$ and shown in 
Fig. \ref{fig:energies} (b) with respect to $g_{B\uparrow}$. 
Indeed $\delta \rho^{(1)}_B(x;g_{B\uparrow}<0)>0$ [Fig. \ref{fig:energies} (b)] in the vicinity of $\rho^{(1)}_I(x;g_{B\uparrow}<0)$ [Fig. \ref{fig:energies} (c)]. 
This density enhancement of the BEC forces the impurities to approach each other leading to the emergence of attractive impurity-impurity induced interactions. 
Similarly for $g_{B\uparrow}<0$ the impurities tend to reside in regions of lower bath density causing a density depletion of the BEC characterized 
by $\delta \rho^{(1)}_B(x;g_{B\uparrow}>0)<0$ [Fig. \ref{fig:energies} (b)]. 
The above-described density depletion of the bath gives rise to the attractive induced interactions analogously to $g_{B\uparrow}<0$. 
It is also worth commenting that for $g_{B\uparrow}>0.5$ $\rho^{(1)}_B(x;g_{B\uparrow}>0.5)$ splits into two branches lying at the Thomas-Fermi edges $\pm R_{TF}$ of the BEC 
[see also Fig. \ref{fig:ground} ($b_5$)]. 
At these values of $g_{B\uparrow}$ $\Delta_+^{N_I}(g_{B\uparrow})$ tends to saturate indicating the impurity-BEC phase separation transition.

\section{Quench induced dynamics}\label{sec:quench}

Next, we study the interspecies interaction quenched dynamics for the mass-balanced multicomponent system which is initially prepared in its 
ground state and characterized by $g_{BB}=0.5$ and $g_{B\uparrow}=0$. 
In this case the Thomas-Fermi radius of the BEC is $R_{TF}\approx4.2$ and the impurities are in a superposition of their spin components described by Eq.~(\ref{Eq:2}). 
We mainly analyze the case of two non-interacting ($g_{II}=0$) impurities and briefly discuss the scenario of two weakly interacting impurity atoms in order 
to expose the effect of their mutual interaction in the dynamics. 

To induce the nonequilibrium dynamics we perform at $t=0$ a sudden change from $g_{B\uparrow}=0$ to either attractive [Sec. \ref{sec:attractive}] or 
repulsive [Sec. \ref{sec:repulsive}] finite values of $g_{B\uparrow}$. 
To examine the emergent dynamics we first discuss the time-evolution of the spin polarization (alias contrast) and its spectrum. 
Consequently we discuss the dynamical response of the impurities in terms of their single-particle densities and the corresponding two-body reduced 
density matrix. 
An effective potential picture for the impurities is  constructed in order to provide an intuitive understanding of the quench dynamics.

\subsection{Interpretation of the contrast of two impurities}\label{sec:structure_interpretation}

To examine the quench-induced dynamics of the two spinor bosonic impurities we first determine the time-evolution of the 
total spin polarization (contrast) $|\braket{\hat{\textbf{S}}(t)}|=\sqrt{\braket{\hat{S}_x(t)}^2 + \braket{\hat{S}_y(t)}^2}$ which 
enables us to infer the dressing of the impurities during the dynamics \cite{Cetina}. 
Note that $\braket{\hat{S}_z(t)}=\braket{\hat{S}_z(t=0)}=0$ since $\left[ \hat{S}_z, \hat{H} \right]=0$ 
and the spin operator in the $k$-th direction ($k=x,y,z$) is given by $\hat{S}_k=(1/N_I)\int dx \sum_{ab} \hat{\Psi}^{\dagger}_a (x) \sigma^k_{ab} \hat{\Psi}_b (x)$, 
with $\sigma^k_{ab}$ denoting the Pauli matrices. 
The contrast for a single impurity has been extensively studied \cite{Grusdt_approaches,Grusdt_RG,Shchadilova,Nishida} and it is related to the so-called Ramsey 
response~\cite{Cetina} and therefore the structure factor. 
The time-dependent  overlap between the interacting and the noninteracting 
states is given by 
\begin{equation}
|\braket{\hat{\bm{S}}(t)}|^2=|\braket{\tilde{\Psi}^0_{BI}| e^{i\tilde{E}_0 t/\hbar} e^{-i\hat{\tilde{H}} t/\hbar}|\tilde{\Psi}^0_{BI}}|^2\equiv\abs{S_1(t)}^2, 
\label{eq:structure_single}
\end{equation}
where $\ket{\tilde{\Psi}^0_{BI}}$ is the spatial part of the MB ground state wavefunction of a single impurity with energy $\tilde{E}_0$ when $g_{BI}=0$. 
$\hat{\tilde{H}}=\hat{P}\hat{H}\hat{P}$ with $\hat{P}$ being the projector operator to the spin-$\uparrow$ configuration, and $\hat{H}$ 
denotes the postquench Hamiltonian [Eq.~(\ref{Htot_system})]. 
Note also that the contrast is chosen here to take values in the interval $[0,1]$. 
From Eq.~(\ref{eq:structure_single}) zero contrast implies that the overlap between the interacting and the non-interacting 
states vanishes signifying an orthogonality catastrophe phenomenon \cite{catstrophe,Nishida}. 
On the other hand, if $|\braket{\hat{\bm{S}}(t)}|^2=1$ then the non-interacting and the interacting states coincide and no quasiparticle is formed. 
Therefore only in the case that $0<|\braket{\hat{\bm{S}}(t)}|^2<1$ we can infer the dressing of the impurity and the formation of a quasiparticle. 

When increasing the number of impurity atoms to $N_I>1$, $|\braket{\hat{\bm{S}}(t)}|^2$ is more complex since 
additional spin states contribute to the MB wavefunction (see Eq.~(\ref{Eq:2})). 
To understand the interpretation of $|\braket{\hat{\bm{S}}(t)}|^2$ during the dynamics we therefore first discuss it 
for the case of two impurities. 
The contrast of two pseudospin-$1/2$ bosonic impurities reads 
\begin{equation}
|\braket{\hat{\bm{S}}(t)}|^2=\frac{1}{4} \abs{A(\ket{1,0};\ket{1,-1})+A^*(\ket{1,0};\ket{1,1})}^2,\label{eq:structure_two_int_imp}
\end{equation}
where the spatial overlap between two different spin configurations namely $\ket{S,S_z}$ and $\ket{S',S'_z}$ is defined as \cite{Tannoudji} 
\begin{equation}
\begin{split}
&A(\ket{S,S_z};\ket{S',S_z'})\equiv\big[\bra{S,S_z}\bra{\Psi_{BI}^0}\big]e^{i\hat{H}t/\hbar}\ket{S,S_z}\\&~~~~~~~~~~~\times\bra{S',S_z'}e^{-i\hat{H}t/\hbar}\big[\ket{\Psi_{BI}^0}\ket{S',S_z'}\big]
\\&=\int dx^{N_B} dx^{N_I} \Psi_{S,S_z}^*(\vec{x}^B,\vec{x}^I;t)\Psi_{S',S_z'}(\vec{x}^B,\vec{x}^I;t), \label{eq:general_structure_factor}
\end{split}
\end{equation}
with $\Psi_{S,S_z}(\vec{x}^B,\vec{x}^I;t)=\frac{\bra{\vec{x}^B,\vec{x}^S}\braket{S,S_z|\Psi(t)}}{\abs{ \abs{\braket{S,S_z|\Psi(0)}}}^2}$ referring to the 
spatial wavefunction corresponding to the spin configuration $\ket{S,S_z}$ and 
$\ket{\Psi_{BI}^0}$ being the spatial part of the initial MB state for two impurities. 
In particular in our case we consider two pseudospin-$1/2$ bosons where $\ket{1,1}\equiv \ket{\uparrow}_1\otimes \ket{\uparrow}_2$, 
$\ket{1,-1}\equiv \ket{\downarrow}_1\otimes \ket{\downarrow}_2$, $\ket{1,0}\equiv \frac{\ket{\uparrow}_1\otimes \ket{\downarrow}_2+\ket{\downarrow}_1\otimes \ket{\uparrow}_2}{\sqrt{2}}$. 
The relevant overlaps read 
$A(\ket{1,0};\ket{1,-1})=e^{-iE_0t/\hbar}\int dx^{N_B} dx^{N_I}  \Psi_{1,0}^*(\vec{x}^B,\vec{x}^I;t) \Psi_{BI}^0 (\vec{x}^B,\vec{x}^I;0)$ and 
$A(\ket{1,0};\ket{1,1})=\int dx^{N_B} dx^{N_I}  \Psi_{1,0}^*(\vec{x}^B,\vec{x}^I;t) \Psi_{1,1}(\vec{x}^B,\vec{x}^I;t)$. 
Recall that a quasiparticle is a free particle that is dressed by the excitations of a bosonic bath via their mutual interactions. 
As a consequence, $\Psi_{BI}^0(\vec{x}^B,\vec{x}^I)$ refers to the wavefunction where no polaron quasiparticle exists since it is the ground state 
wavefunction of the system with $g_{B\uparrow}=0$. 
Moreover, $\Psi_{1,0}(\vec{x}^B;\vec{x}^I)$ and $\Psi_{1,1}(\vec{x}^B;\vec{x}^I)$ denote the wavefunctions where a single and two impurities respectively 
interact with the bosonic gas and therefore describe the formation of a single and two polarons, respectively. 
Accordingly, $A(\ket{1,0};\ket{1,-1})$ provides the overlap between the state of a single and no impurities interacting with the bath, 
while $A(\ket{1,0};\ket{1,1})$ is the overlap between a single and two impurities interacting with the bath. 

As a result, $|\braket{\hat{\bm{S}}(t)}|^2=1$ means that $A(\ket{1,0};\ket{1,-1})=A(\ket{1,0},\ket{1,1})=e^{i\varphi}$ where $\varphi$ is a phase factor. 
The fact that $\abs{A(\ket{1,0};\ket{1,-1})}=1$ implies that the spatial state of a single impurity interacting with the bath is the same as the non-interacting 
one, except for a possible phase factor, and therefore a quasiparticle is not formed. 
Moreover since also $\abs{A(\ket{1,0},\ket{1,1})}=1$ it holds that the state of a single pseudospin-$\uparrow$ interacting impurity coincides with the state 
of two pseudospin-$\uparrow$ impurities interacting with the bath and as 
a consequence with a bare particle due to $\abs{A(\ket{1,0};\ket{1,-1})}=1$. 
Thus, $|\braket{\hat{\bm{S}}(t)}|^2=1$ implies that there is no quasiparticle formation. 
On the contrary for $|\braket{\hat{\bm{S}}(t)}|^2=0$ either $A(\ket{1,0};\ket{1,-1})=A(\ket{1,0},\ket{1,1})=0$ or $A(\ket{1,0};\ket{1,-1})=-A^*(\ket{1,0},\ket{1,1})$ 
should be satisfied. 
In the former case we can deduce the occurrence of an orthogonality catastrophe phenomenon as in the single impurity case while the 
latter scenario is given by the destructive interference of the $A(\ket{1,0};\ket{1,-1})$ and $A(\ket{1,0},\ket{1,1})$ terms. 
However, for $0<|\braket{\hat{\bm{S}}(t)}|^2<1$ the corresponding overlaps acquire finite values and a quasiparticle can be formed. 

Notice also that in the special case of $g_{\uparrow\downarrow}=0$ and $g_{\downarrow\downarrow}=0$ (but $g_{\uparrow\uparrow}$ arbitrary) it can be shown that 
$A(\ket{1,0};\ket{1,-1})=\braket{\tilde{\Psi}_{BI}^0|e^{i\hat{H}t/\hbar}e^{i\tilde{E}_0t/\hbar}|\tilde{\Psi}_{BI}^0}\equiv S_1(t)$. 
The latter is exactly the contrast or the structure factor of a single impurity [Eq. (\ref{eq:structure_single})]. 
Indeed $\ket{\Psi_{BI}^0}=\ket{\tilde{\Psi}_{BI}^0}\otimes\ket{{\psi}_{I}^0}$ for $g_{\downarrow\downarrow}=0$ holds where $\ket{\psi_I^0}$ 
is the single-particle ground state of the impurity while $\ket{\tilde{\Psi}_{BI}^0}$ and $\ket{\Psi_{BI}^0}$ refer to the spatial part of the 
MB ground state wavefunction of a single (energy $\tilde{E}_0$) and two impurities (energy $E_0$) respectively. 
Additionally $\hat{H}$ is the postquench Hamiltonian given by Eq.~(\ref{Htot_system}). 
Consequently, the contrast in this special case acquires the simplified form 
\begin{equation}
|\braket{\hat{\bm{S}}(t)}|^2=\frac{1}{4}\abs{S_1(t)+A^*(\ket{1,0};\ket{1,1})}^2. \label{structure_two_non_int_imp}  
\end{equation} 
Evidently, here $|\braket{\hat{\bm{S}}(t)}|$ depends explicitly on the structure factor $S_1(t)$ of a single impurity allowing 
for a direct interpretation of the dynamical dressing of the two impurities with respect to the single impurity case 
discussed in Ref. \cite{Mistakidis_orth_cat}. 
In the following, $g_{\uparrow\uparrow}=g_{\downarrow\downarrow}=g_{\uparrow\downarrow}\equiv g_{II}$ and as 
a consequence $g_{\uparrow\downarrow}=0$, $g_{\downarrow\downarrow}=0$ is encountered for $g_{II}=0$ while the general case of 
Eq. (\ref{eq:general_structure_factor}) applies for the case of $g_{II}=0.2$ analyzed below.

\subsection{Evolution of the contrast}\label{sec:structure_evolution}

\begin{figure*}[ht]
  	\includegraphics[width=0.78\textwidth]{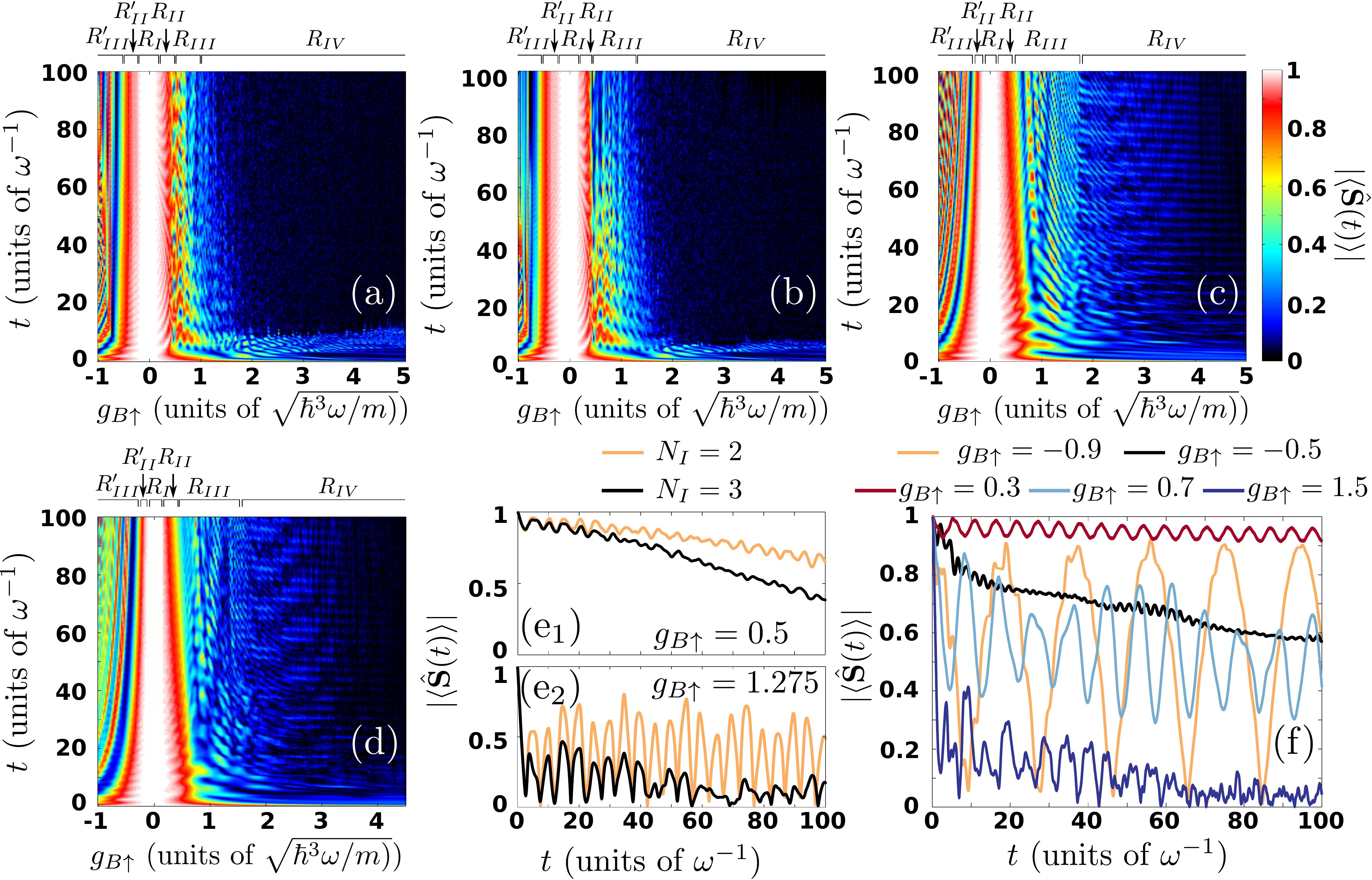}
  	\caption{Time-evolution of the contrast, $|\braket{\hat{\bm{S}}(t)}|$,
  	of two (a) non-interacting ($g_{II}=0$) and (b) weakly 
  	repulsive ($g_{II}=0.2$) impurities immersed in a bath of $N_B=100$ atoms 
  	for different interspecies interaction strengths $g_{B\uparrow}$. 
  	(c) The same as (a) but when considering a few-body bath of $N_B=10$ bosons. 
  	(d) $|\braket{\hat{\bm{S}}(t)}|$ for $N_I=3$ non-interacting impurities inside a few-body bath 
  	consisting of $N_B=10$ atoms. 
  	(e$_1$), (e$_2$) $|\braket{\hat{\bm{S}}(t)}|$ of two non-interacting impurities in a bath of $N_B=10$ bosons for different $g_{B\uparrow}$ (see legends). 
  	(f) Dynamics of $|\braket{\hat{\bm{S}}(t)}|$ for specific postquench interaction strengths (see 
  	legend) when $N_I=2$, $g_{II}=0$ and $N_B=100$. 
  	In all cases the multicomponent system is harmonically trapped and it is initialized in its 
  	ground state with $g_{BB}=0.5$ and $\omega=1$.}
  	\label{fig:contrast} 
\end{figure*}

The dynamics of the two particle contrast $|\braket{\hat{\bm{S}}(t)}|$ is presented in Figs. \ref{fig:contrast} (a)-(c) for both attractive and repulsive postquench interspecies 
interactions $g_{B\uparrow}$. 
In particular, $|\braket{\hat{\bm{S}}(t)}|$ is shown for either two non-interacting [Fig. \ref{fig:contrast} (a)] or interacting [Fig. \ref{fig:contrast} (b)] impurities and $N_B=100$ 
as well as for a few-body bosonic gas with $N_B=10$ and $g_{II}=0$ [Fig. \ref{fig:contrast} (c)]. 
In all cases, six different dynamical regions with respect to $g_{B\uparrow}$ can be identified marked as $R_I$, $R_{II}$, $R_{III}$, $R_{IV}$, $R'_{II}$ and $R'_{III}$. 
Focusing on the system with $N_B=100$ and $g_{II}=0$ these regions correspond to $-0.2\leq g^{R_{I}}_{B\uparrow} < 0.2 $, $0.2\leq g^{R_{II}}_{B\uparrow} <0.4$, 
$0.4\leq g^{R_{III}}_{B\uparrow} <1 $, $1\leq g^{R_{IV}}_{B\uparrow} <5 $, $-0.5\leq g^{R'_{II}}_{B\uparrow} < -0.2 $ and $-1\leq g^{R'_{III}}_{B\uparrow} <-0.5$ respectively [Fig. \ref{fig:contrast} (a)]. 
Specifically, within the very weakly interacting region $R_I$ the contrast is essentially unperturbed remaining unity in the course of the time-evolution and therefore there is no 
quasiparticle formation. 
For postquench interactions lying within $R_{II}$ or $R'_{II}$ the contrast performs small and constant 
amplitude oscillations, weakly deviating from $|\braket{\hat{\bm{S}}(t=0)}|=1$ [Fig. \ref{fig:contrast} (f)]. 
This behavior indicates the generation of two long-lived coherent quasiparticles (see also section \ref{sec:spectrum}). 
Entering the intermediate repulsive interaction region $R_{III}$, $|\braket{\hat{\bm{S}}(t)}|$ exhibits large amplitude ($0<|\braket{\hat{\bm{S}}(t)}|<1$) multifrequency 
temporal oscillations [Fig. \ref{fig:contrast} (f)]. 
The latter signifies the dynamical formation of two Bose polarons which are 
coupled with higher-order excitations of the bosonic bath when compared to regions $R_{II}$ and $R'_{II}$ as we shall expose in section \ref{sec:density_repulsive}. 
For intermediate attractive interactions (region $R'_{III}$) $|\braket{\hat{\bm{S}}(t)}|$ undergoes large amplitude oscillations taking values 
in the interval $0<|\braket{\hat{\bm{S}}(t)}|<1$ [Fig. \ref{fig:contrast} (f)]. 
This response of $|\braket{\hat{\bm{S}}(t)}|$
again signals quasiparticle formation. 
However, in addition to this dynamical dressing the destructive ($|\braket{\hat{\bm{S}}(t)}|=0$) and the 
constructive ($|\braket{\hat{\bm{S}}(t)}|\approx 1$) interference between the states of a single and two Bose polarons can be seen 
(see also Eq. (\ref{eq:structure_two_int_imp}) and its interpretation in section \ref{sec:structure_interpretation}). 

For strong repulsive interactions lying within $R_{IV}$ the contrast shows a fastly decaying amplitude at short evolution times ($0<t<2$) and subsequently fluctuates 
around zero [Fig. \ref{fig:contrast} (f)]. 
This latter behavior of $|\braket{\hat{\bm{S}}(t)}|^2\to 0$ is a manifestation of an orthogonality catastrophe phenomenon of the spontaneously generated short-lived 
($0<t<2$) Bose polarons. 
It is a consequence of the spatial phase separation between the impurity and the bosonic bath (see also Fig. \ref{fig:one_body_repul} (h) and 
the discussion in section \ref{sec:density_repulsive}), where the impurity prefers to reside at the  edges of the BEC background, see also Fig. \ref{fig:energies} (c). 
Note that this behavior is also supported by the effective potential of the impurities, see Eq. (\ref{eq:eff_pot_stat}). 
Most importantly this process results in an energy transfer from the impurity to the BEC, which 
prohibits the revival of the dynamical state of the impurity to its initial one, 
implying $|\braket{\hat{\bm{S}}(t)}|^2\ll 1$. 
Such a mechanism has been also identified to occur for the case of a single impurity, see Ref.~\cite{Mistakidis_orth_cat}. 

The emergence of the different dynamical regions in the evolution of the contrast holds equally when the size of the bath decreases to $N_B=10$ [Fig. \ref{fig:contrast} (c)]. 
For such a few-body scenario region $R_{II}$, where coherently long-lived quasiparticles are formed, becomes slightly wider, i.e. $0.2\leq g^{R_{II}}_{B\uparrow} <0.6$, 
compared to the $N_B=100$ case. 
The most notable difference between the few and the many particle bath takes place in the intermediate interaction region $R_{III}$.
The latter, occurs now at $0.6\leq g^{R_{III}}_{B\uparrow} <1.8 $, 
with $|\braket{\hat{\bm{S}}(t)}|$ performing large amplitude multifrequency oscillations implying in turn the formation of highly excited polaronic states. 
Note that the amplitude of the oscillations of $|\braket{\hat{\bm{S}}(t)}|$ here is larger 
than in the $N_B=100$ case [Fig. \ref{fig:contrast} (a)]. 
Additionally, we observe that $|\braket{\hat{\bm{S}}(t)}|$ decreases smoothly as $g_{B\uparrow}$ increases, which is in sharp contrast to the $N_B=100$ case.
Recall that such a smooth behavior occurring in the few-body scenario 
has already been identified in our discussion of the ground state 
properties and in particular when inspecting the relative distance 
between the impurities. 
Also, the oscillations of $|\braket{\hat{\bm{S}}(t)}|$ ($0<|\braket{\hat{\bm{S}}(t)}|<1$) for 
intermediate attractive interactions (region $R'_{III}$)
being a consequence of the destructive ($|\braket{\hat{\bm{S}}(t)}|=0$) and constructive 
($|\braket{\hat{\bm{S}}(t)}|\approx 1$) interference between the states of a single 
and two Bose polarons are much more prevalent and regular for $N_B=10$ as compared to the $N_B=100$ case. 
Concluding, we can infer that the overall phenomenology of the dynamical formation of quasiparticles as imprinted in the contrast is 
similar for $N_B=10$ and $N_B=100$. 

To test the effect of the number of impurities on the interaction intervals of quasiparticle formation we also consider the case of $N_I=3$ non-interacting, $g_{II}=0$, 
bosons immersed in a few-body bath of $N_B=10$ atoms. 
The dynamics of the corresponding contrast for this system following a quench from $g_{B\uparrow}=0$ to a finite either attractive or repulsive $g_{B\uparrow}$ 
is illustrated in Fig. \ref{fig:contrast} (d). 
As it can be seen, $|\braket{\hat{\bm{S}}(t)}|$ shows a similar behavior to the case of two impurities [Fig. \ref{fig:contrast} (c)] but the regions of finite contrast become 
narrower. 
Particularly, the intermediate repulsive interaction region here occurs for $0.5\leq g^{R_{III}}_{B\uparrow} <1.5 $ instead of 
$0.6\leq g^{R_{III}}_{B\uparrow} <1.8 $ for $N_I=2$. 
Additionally, $|\braket{\hat{\bm{S}}(t)}|$ acquires lower values within the regions $R_{III}$ and $R'_{III}$ for more impurities. 
Moreover, for $N_I=3$ within $R'_{III}$ we observe a pronounced dephasing of the contrast which is absent for the $N_I=2$ case, see Figs. \ref{fig:contrast} (e$_1$), (e$_2$). 
As a consequence, we can deduce that the basic characteristics of the regions of dynamical polaron formation do not significantly 
change for a larger number of impurities in the regime $N_I\ll N_B$. 

Finally, we discuss $|\braket{\hat{\bm{S}}(t)}|$ for weakly interacting impurities. 
Comparing the temporal evolution of $|\braket{\hat{\bm{S}}(t)}|$ for $g_{II}=0.2$ [Fig. \ref{fig:contrast} (b)] to the one for $g_{II}=0$ [Fig. \ref{fig:contrast} (a)] 
we observe that the extent of the above-described dynamical regions ($R_I$, $R_{II}$, $R_{III}$, $R_{IV}$, $R'_{II}$ and $R'_{III}$) can be tuned via $g_{II}$. 
For instance, region $R_{II}$ occurs at $0.2\leq g^{R_{II}}_{B\uparrow} <0.4$ for $g_{II}=0.2$ instead of $0.2\leq g^{R_{II}}_{B\uparrow} <0.5$ when $g_{II}=0$, 
while region $R_{III}$ takes place at $0.4\leq g^{R_{III}}_{B\uparrow} <1.3$ if $g_{II}=0.2$ and within $0.5\leq g^{R_{III}}_{B\uparrow} <1$ in the non-interacting scenario. 
Also region $R_{IV}$ where the orthogonality catastrophe takes place is shifted to slightly larger interactions for $g_{II}=0.2$ compared to the $g_{II}=0$ case. 
Interestingly we observe that the contrast within $R_{III}$ and $R'_{III}$ exhibits a decaying tendency for long evolution times $t>50$ in the presence of weak impurity-impurity 
interactions, a behavior which is absent when $g_{II}=0$. 

\begin{figure*}[ht]
  	\includegraphics[width=0.78\textwidth]{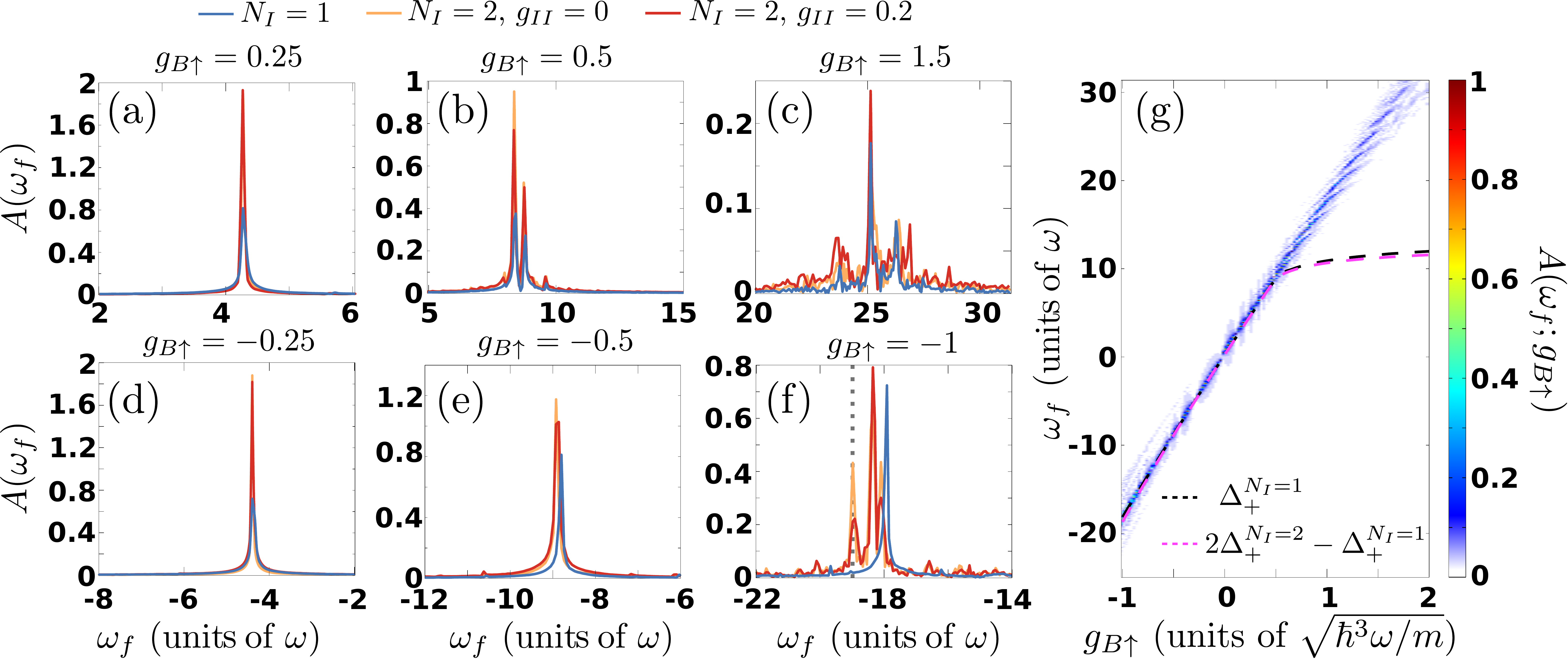}
  	\caption{Excitation spectrum, $A(\omega_f)$, of a single, two non-interacting, 
  	and two interacting bosonic impurities (see legend) for different 
  	interspecies interaction strengths $g_{BI}$. 
  	Note that for better visibility $A(\omega_f)$ for $N_I=2$ is scaled by a factor of two when compared to the $N_I=1$ case. 
  	The dashed line in Fig. \ref{fig:spectrum} (f) indicates the position of the two polaron resonance i.e. $2\Delta_+^{N_I=2}-\Delta_+^{N_I=1}=-18.98$. 
  	(g) $A(\omega_f)$ of two non-interacting impurities with varying $g_{BI}$. 
  	The dashed lines indicate the expected position of the polaronic resonances $\Delta_+^{N_I}(g_{B\uparrow})$ (see legend). 
  	The harmonically trapped bosonic mixture is initialized in its ground state and consists of 
  	$N_B=100$ atoms with $g_{BB}=0.5$ and either $N_I=1$ or $N_I=2$ impurities.}
  	\label{fig:spectrum} 
\end{figure*}

\subsection{Spectrum of the contrast}\label{sec:spectrum}

To quantify the excitation spectrum of the impurity we calculate the spectrum of the contrast, namely 
\begin{equation}
A(\omega_f)=\frac{1}{\pi} \left|\int^{\infty}_0 dt~e^{i \omega_f t}~|\braket{\hat{\bm{S}}(t)}|e^{i \tan^{-1} \frac{\braket{\hat S_x(t)}}{\braket{\hat S_y(t)}}} \right|.
\end{equation}
Recall that at low impurity densities and weak interspecies interactions it has been shown that $|\braket{\hat{\bm{S}}(t)}|$ is 
proportional to the so-called spectral function of quasiparticles \cite{Cetina,Parishsuper,Nishida}. 
Figure \ref{fig:spectrum} presents $A(\omega_f)$ in the case of a single and two either non-interacting ($g_{II}=0$) or weakly 
interacting ($g_{II}=0.2$) impurities when $N_B=100$ for different interspecies couplings of either sign. 
Evidently, for weak $g_{B\uparrow}$ belonging either to region $R_{II}$ with $g_{B\uparrow}=0.25$ [Fig. \ref{fig:spectrum} (a)] 
or $R'_{II}$ with $g_{B\uparrow}=-0.25$ [Fig. \ref{fig:spectrum} (d)] we observe a single peak in $A(\omega_f)$ located at $\omega_f\approx4.27$ and $\omega_f\approx-4.39$ respectively.   
This single peak occurs independently of the number of impurities and their intraspecies interactions. 
Therefore, this peak at small $g_{B\uparrow}=\pm0.25$ corresponds to the long-time evolution of a well-defined repulsive or attractive Bose 
polaron respectively. 
Within region $R_{III}$ e.g. at $g_{B\uparrow}=0.5$ two dominant peaks occur in $A(\omega_f)$ [Fig. \ref{fig:spectrum} (b)] at frequencies 
$\omega_f\approx8.42$ and $\omega_f\approx8.79$ for both the $N_I=1$ and $N_I=2$ cases. 
Accordingly, these two peaks suggest the formation of a quasiparticle dressed, for higher frequencies, by higher-order excitations of the 
BEC background. 

Entering the strongly interspecies repulsive region $R_{IV}$ a multitude of frequencies are imprinted in the impurity's excitation spectrum 
e.g. at $g_{B\uparrow}=1.5$, see Fig. \ref{fig:spectrum} (c). 
The number of the emerging frequencies is larger for the two compared to the single impurity but does not significantly depend on $g_{II}$ for $N_I=2$. 
For instance, when $N_I=1$ mainly three predominant peaks centered at $\omega_f\approx23.75$, $\omega_f\approx25.13$, and $\omega_f\approx26.26$ appear in $A(\omega_f)$ whilst 
for $N_{I}=2$ and $g_{II}=0$ five dominantly contributing frequencies located at $\omega_f\approx22.31$, $\omega_f\approx23.81$, $\omega_f\approx25.2$, $\omega_f\approx26.39$ 
and $\omega_f\approx27.52$ occur. 
These frequency peaks correspond to even higher excited states of the quasiparticle than the ones within the region $R_{III}$. 
We note that for values of $g_{B\uparrow}$ deeper in $R_{IV}$ a variety of low amplitude but large valued frequency peaks occur in $A(\omega_f)$. 
This fact indicates that the impurities tend to populate a multitude of states indicating the manifestation of the polaron orthogonality catastrophe 
as discussed in Refs. \cite{Mistakidis_orth_cat,Mistakidis_inject_imp} (results not shown here). 

Turning to intermediate attractive interactions lying within $R'_{III}$ such as $g_{B\uparrow}=-0.5$ a single frequency peak can be seen in $A(\omega_f)$ whose 
frequency is shifted towards more negative values for $N_I=2$ compared to $N_I=1$ and also for increasing $g_{II}$ [Fig. \ref{fig:spectrum} (e)]. 
Specifically, when $N_I=1$ the aforementioned peak occurs at $\omega_f\approx-8.79$ while for $N_{I}=2$ and $g_{II}=0$ [$g_{II}=0.2$] it lies at $\omega_f\approx-8.92$ [$\omega_f\approx-8.86$]. 
This peak indicates the generation of an attractive Bose polaron. 
A further increase of the attraction, e.g. $g_{B\uparrow}=-1$, leads to the appearance of three quasiparticle peaks in $A(\omega_f)$ when $N_I=2$ and either $g_{II}=0$ or $g_{II}=0.2$, 
centered at $\omega_f\approx-18.1$, $\omega_f\approx-18.35$ and $\omega_f\approx-18.98$, but only one for $N_I=1$ with $\omega_f\approx-17.91$, as shown in Fig. \ref{fig:spectrum} (f). 
This change of $A(\omega_f)$ for increasing $N_I$ within the regions $R'_{II}$ and $R'_{III}$ demonstrates the prominent role of induced interactions for 
attractive interspecies ones. 
More specifically for $N_I=2$, $A(\omega_f)$ possesses additional quasiparticle peaks as compared to the $N_I=1$ case. 
Indeed, according to Eq. (\ref{eq:structure_two_int_imp}) we can predict at least two peaks at positions $\omega_f=\Delta_+^{N_I=1}=-17.96$ and 
$\omega_f=2\Delta_+^{N_I=2}-\Delta_+^{N_I=1}=-18.98$ explaining 
two of the above indentified peaks. 
The third dominant peak at $\omega_f=18.35$ appearing in the spectrum is attributed to the occupation of an excited state with $S_z=1$ [see also Eq. (\ref{Eq:2})] 
according to Eq. (\ref{eq:structure_two_int_imp}). 
Recall that the $\ket{1,1}$ spin state in the time-evolved wavefunction [Eq. (\ref{Eq:2})]
corresponds to the two polaron case while $\ket{1,0}$ contains only one polaron and the 
$\ket{1,-1}$ describes impurities that do not interact with the bath and thus no polarons. 
The aforementioned population of the additional polaronic states for $N_I=2$ is a clear evidence of impurity-impurity induced interactions. 

The overall behavior of the excitation spectrum $A(\omega_f;g_{B\uparrow})$ for $N_I=2$ and $g_{II}=0$ is shown in Fig. \ref{fig:spectrum} (g) with varying $g_{B\uparrow}$. 
Evidently, the position of the dominant quasiparticle peak in terms of $\omega_f$ increases almost linearly for larger $g_{B\uparrow}$. 
This behavior essentially reflects the linear increase of the energy of the initial state $\ket{\Psi(0)}$ [Eq. (\ref{Eq:2})] directly after the quench. 
Moreover, comparing the position of the dominant quasiparticle peak with $\Delta_+^{N_I=1}$ reveals that for $g_{B\uparrow}>0.5$, while the latter saturates, 
the former increases and additional peaks appear in the spectrum $A(\omega_f;g_{B\uparrow})$. 
These peaks correspond to excited states of the system and already for $g_{B\uparrow}>1$ the ground
states corresponding to $\Delta^{N_I}_{+}$ cease to be populated during the dynamics. 
In a similar fashion, such additional quasiparticle peaks occur also for attractive interactions, see Fig. \ref{fig:spectrum} (g) for $g_{B\uparrow}<-0.5$. 
In this case the additional quasiparticle peaks stem from the induced interactions resulting in the 
presence of a peak at $\omega_f=2\Delta_+^{N_I=2}-\Delta_+^{N_I=1}\neq \Delta_+^{N_I=1}$ 
and other ones which correspond to the occupation of higher-lying excited polaronic states with
$S_z=1$ [Eq. (\ref{Eq:2})]. 
Note that such an almost linear behavior of the polaronic spectrum is reminiscent of the corresponding three-dimensional scenario but away from the 
Feshbach resonance regime. 
The latter corresponds in one-dimension to an interspecies Tonks-Girardeau interaction regime which is not addressed in the present work. 
We remark that in one-dimension there is no molecular bound state occuring for repulsive interactions. 

Summarizing, we can infer that the quasiparticle excitation spectrum depends strongly on the value of the postquench interspecies interaction strength and 
also on the number of impurities outside the weakly attractive and repulsive coupling regimes \cite{Parishsuper}. 
However, this behavior is also slightly altered when going from two non-interacting to two weakly interacting impurities. 
For a relevant discussion on the lifetime of the above-described spectral features we refer the interested reader to Ref. \cite{Mistakidis_pump}. 
It is also important to mention that in the weakly interacting impurity-BEC regime where the contrast is finite in the course of the evolution the spectral 
function $A(\omega_f)$ corresponds to the injection spectrum in the framework of the reverse rf spectroscopy \cite{Rath_approaches,Schmidt_rev}. 
\begin{figure*}[ht]
  	\includegraphics[width=0.9\textwidth]{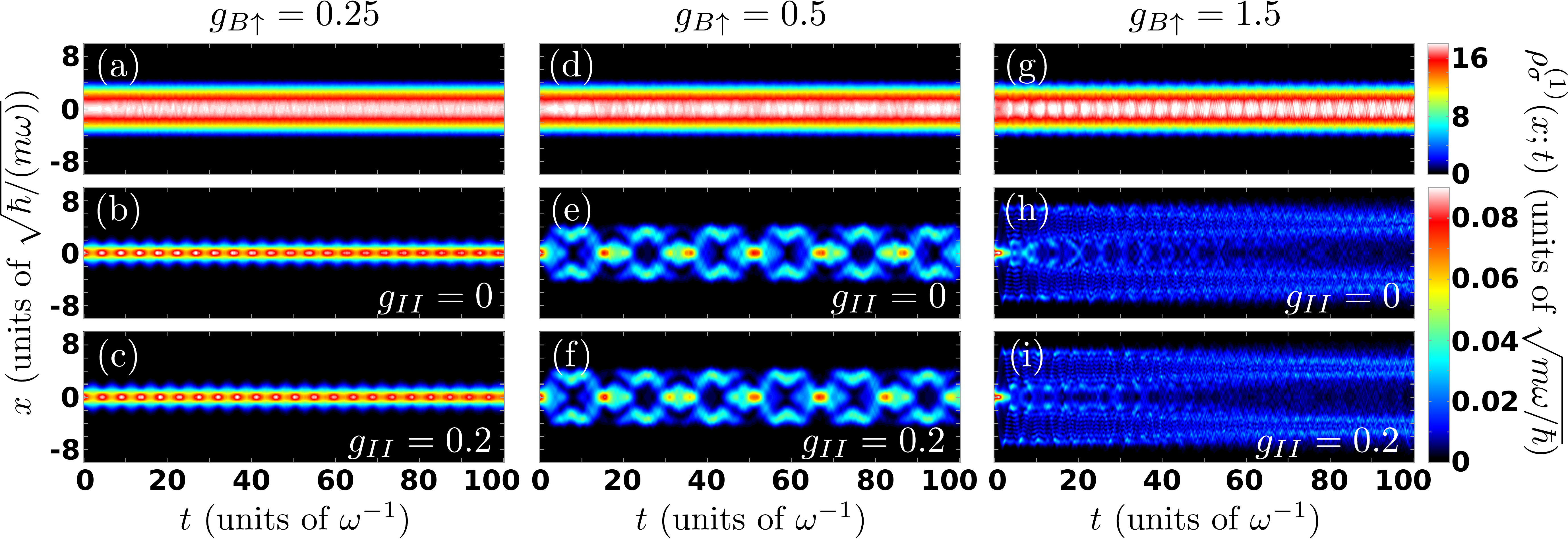}
  	\caption{Time-evolution of the single-particle density, $\rho^{(1)}_{\sigma}(x;t)$, of (a), 
  	(d), (g) the bosonic bath ($\sigma=B$) and (b), (e), (h) the pseudospin-$\uparrow$ part 
  	($\sigma=\uparrow$) of the 
  	two non-interacting impurities for different postquench interspecies repulsions $g_{B\uparrow}$ 
  	(see legend). 
  	Evolution of $\rho^{(1)}_{\uparrow}(x;t)$ for two weakly interacting, $g_{II}=0.2$, impurities 
  	following a quench to (c) $g_{B\uparrow}=0.25$, (f) $g_{B\uparrow}=0.5$ and 
  	(i) $g_{B\uparrow}=1.5$. 
  	The Bose-Bose mixture consists of $N_B=100$ atoms and $N_I=2$ impurities with $g_{BB}=0.5$ and 
  	it is trapped in a harmonic oscillator potential.}
  	\label{fig:one_body_repul} 
\end{figure*}

\subsection{Quench to repulsive interactions}\label{sec:repulsive}

Below we further analyze the dynamical response of the multicomponent system, and especially of the impurities, following an interspecies 
interaction quench from $g_{B\uparrow}=0$ to $g_{B\uparrow}>0$ within the above identified dynamical regions of the contrast. 
In particular, we explore the dynamics of the system on both the single- and the two-body level 
and further develop an effective potential picture to 
provide a more concrete interpretation of the emergent phenomena. 
We mainly focus on the nonequilibrium dynamics of two non-interacting impurities ($g_{II}=0$) and subsequently discuss whether possible alterations 
might occur for weakly interacting ($g_{II}=0.2$) impurities. 
Also, in the following, only the temporal-evolution of the pseudospin-$\uparrow$ part of the impurities is discussed since the pseudospin-$\downarrow$ 
component does not interact with the bosonic medium.

\subsubsection{Density evolution and effective potential}\label{sec:density_repulsive}

\begin{figure*}[ht]
  	\includegraphics[width=0.7\textwidth]{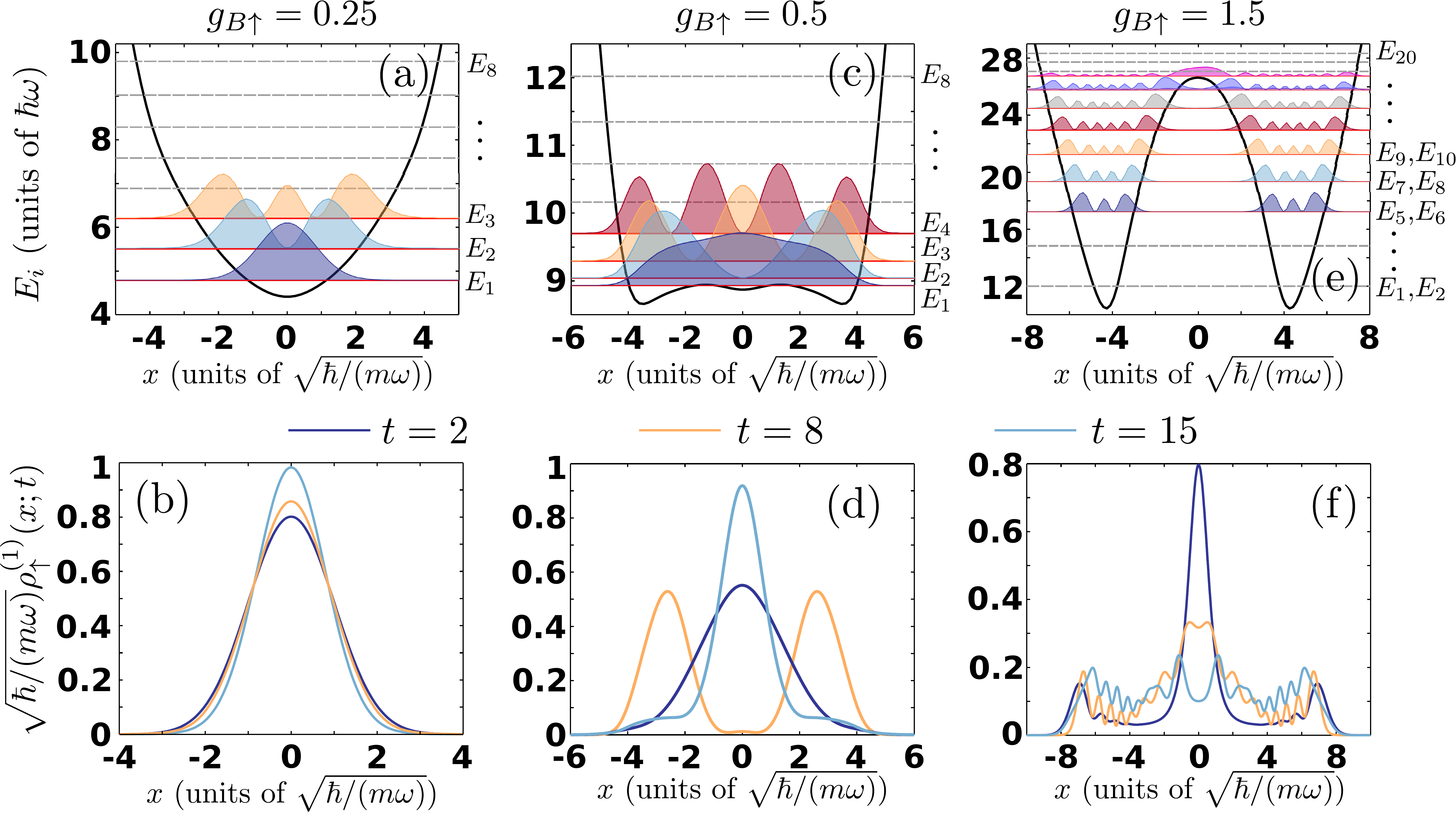}
  	\caption{Time-averaged effective potential, $\bar{V}_I^{eff}(x)$, over $T=100$ 
  	[Eq. (\ref{effective_potential_repulsive})] of the impurities for (a) weak 
  	$g_{B\uparrow}=0.25$, (c) intermediate $g_{B\uparrow}=0.5$ and (e) 
  	strong $g_{B\uparrow}=1.5$ interspecies repulsions.  
  	The densities of the single-particle eigenstates and eigenenergies $E_i$, $i=1,2,\dots$ 
  	of $\bar{V}_I^{eff}(x)$ are also shown. 
  	Profiles of the single-particle density of the two non-interacting impurities 
  	at distinct time-instants of the evolution following an interspecies interaction quench to 
  	(b) $g_{B\uparrow}=0.25$, (d) $g_{B\uparrow}=0.5$ and (f) $g_{B\uparrow}=1.5$ obtained within 
  	the MB approach.}
  	\label{fig:effective_pot_repulsive} 
\end{figure*}

\begin{figure*}[ht]
  	\includegraphics[width=0.7\textwidth]{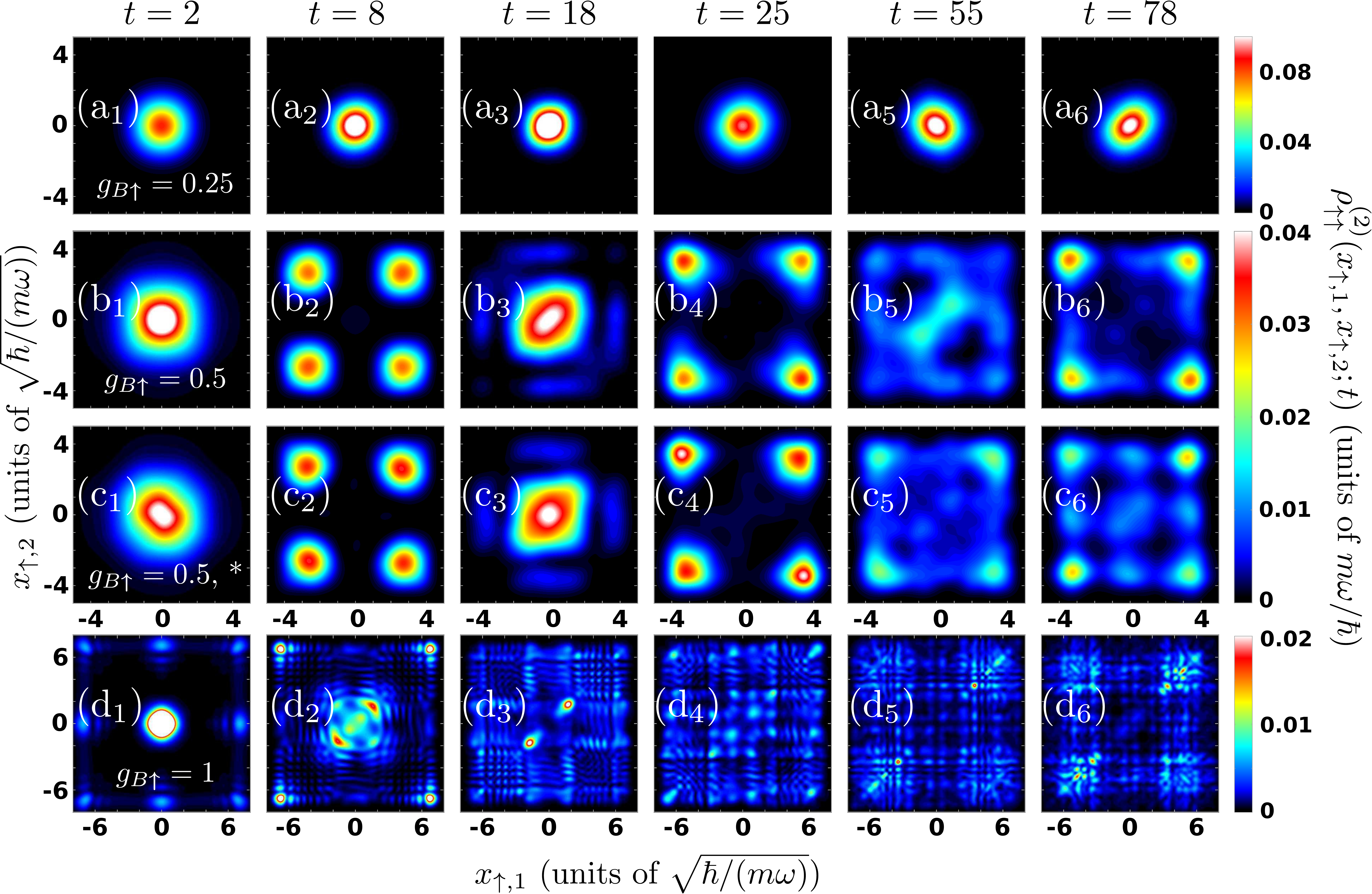}
  	\caption{Two-body reduced density matrix, $\rho^{(2)}_{\uparrow\uparrow}(x_1,x_2;t)$, between 
  	the two pseudospin-$\uparrow$ non-interacting ($g_{II}=0$) bosonic impurities at different time 
  	instants of the MB evolution (see legend) 
  	following an interspecies interaction quench to ($a_1$)-($a_6$) $g_{B\uparrow}=0.25$, ($b_1$)-
  	($b_6$) $g_{B\uparrow}=0.5$ and ($d_1$)-($d_6$) $g_{B\uparrow}=1.5$. 
  	($c_1$)-($c_6$) The same as in ($b_1$)-($b_6$) but for two weakly interacting $g_{II}=0.2$ 
  	impurities. 
  	The harmonically trapped bosonic mixture is composed by $N_B=100$ atoms with $g_{BB}=0.5$ and 
  	$N_I=2$ impurities and it is initialized in its corresponding 
  	ground state configuration. }
  	\label{fig:two_body_repulsive} 
\end{figure*}

To visualize the spatially resolved dynamics of the system on the single-particle level we first inspect the time-evolution 
of the $\sigma$-species single-particle density $\rho^{(1)}_{\sigma}(x;t)$ [Eq. (\ref{eq:single_particle_density_matrix})] illustrated in Fig. \ref{fig:one_body_repul}. 
For weak postquench interspecies repulsions lying within the region $R_{II}$ e.g. $g_{B\uparrow}=0.25$, such that $g_{B\uparrow}<g_{BB}$, the impurities [see Fig. \ref{fig:one_body_repul} (b)] 
exhibit a breathing motion of frequency $\omega_{br}^I\approx 1.44$ inside the bosonic medium \cite{Sartori,Hannes}. 
Moreover, at initial evolution times ($t<60$) the amplitude of the breathing is almost constant whilst later on ($t>60$) it shows a slightly decaying tendency, see for instance 
the smaller height of the density peak at $t=70$ compared to $t=20$ in Fig. \ref{fig:one_body_repul} (b). 
This decaying amplitude can be attributed to the build up of impurity-impurity correlations in the course of the evolution \cite{Mistakidis_two_imp_ferm} due to the 
presence of induced interactions discussed later on, see also Fig. \ref{fig:two_body_repulsive} (a). 
The breathing motion of the impurities is directly captured by the periodic contraction and expansion in the shape of the instantaneous density profiles 
of $\rho^{(1)}_{\uparrow}(x;t)$ depicted in Fig. \ref{fig:effective_pot_repulsive} (b). 
On the other hand, the bosonic gas remains essentially unperturbed [Fig. \ref{fig:one_body_repul} (a)] throughout the dynamics, showing only tiny distortions 
from its original Thomas-Fermi cloud due to its interaction with the impurity. 

An intuitive understanding of the observed dynamics of the impurities is provided with the aid of an effective potential picture. 
Indeed, the impurity-BEC interactions can be taken into account, to a very good approximation, by employing a modified external potential for the impurities. 
The latter corresponds to the time-averaged effective potential created by the 
harmonic oscillator and the density of the bosonic gas \cite{Mistakidis_orth_cat,Hannes,Mistakidis_inject_imp,Theel} namely 
\begin{equation}
 \bar{V}_I^{eff}(x)=\frac{1}{2}m\omega^2x^2+\frac{g_{BI}}{T}\int_0^T dt  \rho_B^{(1)}(x;t). \label{effective_potential_repulsive}
\end{equation} 
The averaging process aims to eliminate the emergent very weak distortions on the instantaneous density of the BEC 
$\rho^{(1)}_B(x;t)$, and it is performed herein over $T=100$. 
These distortions being a consequence of the motion of the impurities within the BEC are imprinted as a slow and very weak amplitude breathing 
motion of $\rho^{(1)}_B(x;t)$ with $\omega_{br}^B\approx1.82$, hardly visible in Fig. \ref{fig:one_body_repul} (a). 
They are canceled out in our case for $T>20$. 
Note that $\omega_{br}^B<2$ is attributed to the repulsive character of the BEC background 
which negatively shifts its breathing frequency from the corresponding non-interacting value \cite{Schmitz_breath}. 
At $g_{B\uparrow}=0.25$ this $\bar{V}_I^{eff}(x)$ takes the form of a modified harmonic oscillator potential illustrated in Fig. \ref{fig:effective_pot_repulsive} (a) 
together with the densities of its first few single-particle eigenstates. 
Furthermore, assuming the Thomas-Fermi approximation for $\rho^{(1)}_B(x,t)$ the effective trapping frequency of the impurities corresponds to 
$\omega_{eff}=\omega\sqrt{1-\frac{g_{B\uparrow}}{g_{BB}}}$. 
Therefore their expected effective breathing frequency would be $\omega_{br}^{eff,I}=2\omega_{eff}\approx1.41$ which is indeed in a very good agreement with 
the numerically obtained $\omega_{br}^I$. 
The discrepancy between the prediction of the effective potential and the MB approach is attributed to the approximate character of the 
effective potential which does not account for possible correlation induced shifts to the breathing frequency. 
Moreover, in the present case the impurities which undergo a breathing motion within $\bar{V}_I^{eff}(x)$ reside predominantly in its energetically 
lowest-lying state $E_1$, see Fig. \ref{fig:effective_pot_repulsive} (a). 
It is also important to mention that this effective potential approximation is adequate only for weak interspecies interactions where the 
impurity-BEC entanglement is small \cite{Mistakidis_orth_cat,Mistakidis_inject_imp}. 
Note also that the inclusion of the Thomas-Fermi approximation in the effective potential of Eq. (\ref{effective_potential_repulsive}) can not adequately describe 
the impurities dynamics when they reach the edges of the bosonic cloud, see \cite{Mistakidis_eff_mass} for more details. 
However in this case $\rho^{(1)}_{\uparrow}(x;t)$ lies within $\rho^{(1)}_{B}(x;t)$ throughout the evolution indicating the miscible character of the dynamics 
for $g_{B\uparrow}<g_{BB}$ \cite{Mistakidis_orth_cat,mistakidis_phase_sep}. 
Furthermore, for these weak postquench interspecies repulsions a similar to the above-described dynamics takes place also for two weakly ($g_{II}=0.2$) 
repulsively interacting impurities as shown in Fig. \ref{fig:one_body_repul} (c). 
The impurities undergo a breathing motion within the bosonic medium in the course of the time-evolution exhibiting a slightly larger oscillation frequency 
than for the $g_{II}=0$ case but with the same amplitude [hardly visible by comparing Figs. \ref{fig:one_body_repul} (b) and (c)]. 

For larger postquench interaction strengths $g_{B\uparrow}=0.5$ (region $R_{III}$), i.e. close to the intraspecies interaction of the bosonic bath $g_{BB}$, the impurities show a more complex 
dynamics compared to the weak interspecies repulsive case [Fig. \ref{fig:one_body_repul} (e)]. 
Also, the BEC medium performs a larger amplitude breathing motion [Fig.~\ref{fig:one_body_repul} (d)] compared to the $g_{B\uparrow}=0.25$ scenario 
but again with a frequency $\omega_{br}^B\approx 1.82$. 
Focusing on the impurities motion, we observe that at short evolution times ($0<t<5$) after the quench $\rho^{(1)}_{\uparrow}(x;t)$ expands and then splits into 
two counterpropagating density branches with finite momenta that travel towards the edges of the bosonic cloud, see Fig. \ref{fig:one_body_repul} (e) 
and the profiles shown in Fig. \ref{fig:effective_pot_repulsive} (d). 
The appearance of these counterpropagating density branches is a consequence of the interaction quench which imports energy 
into the system. 
Reaching the edges of $\rho^{(1)}_{B}(x;t)$ the density humps of $\rho^{(1)}_{\uparrow}(x;t)$ are reflected back towards the trap center ($x=0$) where they 
collide around $t\approx15$ forming a single density peak [Fig. \ref{fig:effective_pot_repulsive} (d)]. 
The aforementioned impurity motion repeats itself in a periodic manner for
all evolution times [Fig. \ref{fig:one_body_repul} (e)].
Here, the underlying time-averaged effective potential [Eq. (\ref{effective_potential_repulsive})] corresponds to a highly deformed harmonic oscillator 
possessing an almost square-well like profile as illustrated in Fig. \ref{fig:effective_pot_repulsive} (c). 
Moreover, a direct comparison of the densities of the lower-lying single-particle eigenstates of $\bar{V}_I^{eff}(x)$ [Fig. \ref{fig:effective_pot_repulsive} (c)] 
with the density profile snapshots of $\rho^{(1)}_{\uparrow}(x;t)$ of the MB dynamics [Fig. \ref{fig:effective_pot_repulsive} (d)] reveals that the impurities predominantly 
reside in a superposition of the two lower-lying excited states ($E_1$ and $E_2$) of $\bar{V}_I^{eff}(x)$. 
Additionally in the case of two weakly repulsively interacting impurities, shown in Fig. \ref{fig:one_body_repul} (f), the impurities' motion remains qualitatively the same. 
However, due to the inclusion of intraspecies repulsion the impurities possess a slightly larger overall oscillation frequency and the collisional 
patterns at the trap center appear to be modified as compared to the $g_{II}=0$ case.  

Turning to strong postquench repulsions, i.e. $g_{B\uparrow}=1.5\gg g_{BB}$ which belongs to $R_{IV}$, the dynamical response of the impurities is greatly altered 
and the bosonic gas exhibits an enhanced breathing dynamics as compared to the weak and intermediate interspecies repulsions discussed above. 
Initially $\rho^{(1)}_{\uparrow}(x;t=0)$ consists of a density hump located at the trap center which, following the interaction quench, breaks into two 
density fragments, as illustrated in Fig. \ref{fig:one_body_repul} (h), each of them exhibiting a multihump structure [see also Fig. \ref{fig:effective_pot_repulsive} (f)]. 
Note that the density hump at the trap center remains the dominant contribution of $\rho^{(1)}_{\uparrow}(x;t)$ until it eventually fades out for $t>5$, see Fig. \ref{fig:one_body_repul} (h). 
This multihump structure building upon $\rho^{(1)}_{\uparrow}(x;t)$ is clearly captured in the instantaneous density profiles depicted in Fig. \ref{fig:effective_pot_repulsive} (f). 
Remarkably, the emergent impurity density fragments that are symmetrically placed around the trap center ($x=0$) perform a damped oscillatory motion in time around 
the edges of the Thomas-Fermi radius of the bosonic gas, see in particular Figs. \ref{fig:one_body_repul} (g), (h). 

The emergent dynamics of the impurities can also be interpreted to lowest order approximation (i.e. excluding correlation effects) by invoking the 
corresponding effective potential which for these strong interspecies repulsions has the form of the double-well potential shown 
in Fig. \ref{fig:effective_pot_repulsive} (e). 
Comparing the shape of the densities of the eigenstates of $\bar{V}_I^{eff}(x)$ [Fig. \ref{fig:effective_pot_repulsive} (e)] with the density profiles $\rho^{(1)}_{\uparrow}(x;t)$ 
[Fig. \ref{fig:effective_pot_repulsive} (f)] obtained within the MB dynamics simulations it becomes evident that the impurities reside in a superposition of higher-lying states of the effective potential. 
Furthermore the double-well structure of $\bar{V}_I^{eff}(x)$ suggests that each of the observed density fragments of the impurities is essentially trapped in each of the corresponding two sites 
of $\bar{V}_I^{eff}(x)$. 
Of course, as already mentioned above, for these strong interactions $\bar{V}_I^{eff}(x)$ provides only a crude description of the impurity dynamics 
since it does not account for both intra- and interspecies correlations that occur during the MB dynamics. 
However $\bar{V}_I^{eff}(x)$ enables the following intuitive picture for the impurity dynamics. 
Namely, the damped oscillations of $\rho^{(1)}_{\uparrow}(x;t)$ designate that the pseudospin-$\uparrow$ impurities at initial times are in a superposition state 
of a multitude of highly excited states [see e.g. Fig.~\ref{fig:effective_pot_repulsive} (f) at $t=8$] while for later times they reside 
in a superposition of lower excited states [see e.g. Fig. \ref{fig:effective_pot_repulsive} (f) at $t=15$]. 
We should also remark that a similar overall dynamical behavior on the single-particle level has been reported in the case of a single spinor impurity 
and has been also related to an enhanced energy transfer from the impurity to the bosonic bath \cite{Mistakidis_orth_cat,Mistakidis_inject_imp,Nielsen,Lampo}. 
Such an energy transfer process takes place also in the present case (results not shown here). 
Another important feature of the observed dynamical response of the impurities is the fact that they are not significantly affected by the presence of weak 
intraspecies interactions. 
This can be seen by inspecting Fig. \ref{fig:one_body_repul} (i) which shows the time-evolution of $\rho^{(1)}_{\uparrow}(x;t)$ for $g_{II}=0.2$. 
Here, the most noticeable difference when compared to the $g_{II}=0$ scenario is that the splitting of $\rho^{(1)}_{\uparrow}(x;t)$ into two branches occurs at shorter 
time scales [compare Figs. \ref{fig:one_body_repul} (h), (i)] due to the additional intraspecies repulsion.  

\subsubsection{Dynamics of the two-body reduced density matrix}\label{sec:two_body_repulsive}

To investigate the development of impurity-impurity correlations during the quench dynamics we next resort to the time-evolution of the pseudospin-$\uparrow$ impurity 
intraspecies two-body reduced matrix $\rho^{(2)}_{\uparrow\uparrow}(x_1,x_2;t)$ [Eq. (\ref{eq:two_body_density})]. 
Recall that $\rho^{(2)}_{\uparrow\uparrow}(x_1,x_2;t)$ provides the probability of finding at time $t$ a pseudospin-$\uparrow$ boson at location $x_1$ and 
a second one at $x_2$ \cite{mistakidis_phase_sep,Erdmann_phase_sep}. 
Most importantly, it allows us to monitor the two-body spatially resolved dynamics of the impurities and infer whether they move 
independently or correlate with each other \cite{induced_int_artem,Bipolaron,Mistakidis_two_imp_ferm}. 

Figure \ref{fig:two_body_repulsive} shows $\rho^{(2)}_{\uparrow\uparrow}(x_1,x_2;t)$ at specific time-instants of the evolution of two 
non-interacting [Fig. \ref{fig:two_body_repulsive} ($a_1$)-($b_6$) and ($d_1$)-($d_6$)] as well as weakly interacting [Fig. \ref{fig:two_body_repulsive} ($c_1$)-($c_6$)] 
impurities for different postquench interspecies repulsions. 
To reveal the role of induced impurity-impurity correlations via the bath we mainly focus on two initially non-interacting impurities where 
$\rho^{(2)}_{\uparrow\uparrow}(x_1,x_2;t=0)=\rho_{\uparrow}^{(1)}(x_1,t=0)\rho_{\uparrow}^{(1)}(x_2,t=0)/2$ since 
$g_{II}=0$ and initially $g_{B\uparrow}=0$. 
As already discussed in section \ref{sec:density_repulsive} for weak interspecies postquench repulsions, namely $g_{B\uparrow}=0.25$ (region $R_{II}$), the impurities perform a breathing 
motion on the single-particle level [Fig. \ref{fig:one_body_repul} (b)] exhibiting a decaying amplitude for large evolution times. 
Accordingly, inspecting $\rho^{(2)}_{\uparrow\uparrow}(x_1,x_2;t)$ [Figs. \ref{fig:two_body_repulsive} ($a_1$)-($a_6$)] we observe that the impurities 
are likely to reside together close to the trap center since $\rho^{(2)}_{\uparrow\uparrow}(-2<x_1<2,-2<x_2<2;t)$ is mainly populated throughout the evolution. 
In particular, at initial times $\rho^{(2)}_{\uparrow\uparrow}(-2<x_1<2,-2<x_2<2;t)$ shows a Gaussian-like distribution which contracts [Fig. \ref{fig:two_body_repulsive} ($a_2$)] 
and expands [Fig. \ref{fig:two_body_repulsive} ($a_3$), ($a_4$)] during the dynamics as a consequence of the aforementioned breathing motion. 
Deeper in the evolution $\rho^{(1)}_{\uparrow}(x;t)$ decays and $\rho^{(2)}_{\uparrow\uparrow}(x_1,x_2;t)$ is deformed along its diagonal [Figs. \ref{fig:two_body_repulsive} ($a_4$), ($a_6$)] or its 
anti-diagonal [Fig. \ref{fig:two_body_repulsive} ($a_5$)] indicating that the impurities tend to be slightly apart or at the same location respectively. 
This is indicative of the admittedly weak induced interactions as the breathing mode along the anti-diagonal of $\rho^{(2)}_{\uparrow\uparrow}(x_1,x_2;t)$ 
(relative coordinate breathing mode) does not possess exactly the same frequency as the breathing along the diagonal (center-of-mass breathing mode). 

For larger interspecies repulsions e.g. for $g_{B\uparrow}=0.5$ (region $R_{III}$) the two-body dynamics of the impurities is significantly altered, see Figs. \ref{fig:two_body_repulsive} ($b_1$)-($b_6$). 
At the initial stages of the dynamics the impurities reside together in the vicinity of the trap center as $\rho^{(2)}_{\uparrow\uparrow}(-3<x_1<3,-3<x_2<3;t)$ is predominantly populated. 
However for later times two different correlation patterns appear in $\rho^{(2)}_{\uparrow\uparrow}(x_1,x_2;t)$ in a periodic manner. 
Recall that for these interactions $\rho^{(1)}_{\uparrow}(x;t)$ splits into two counterpropagating density branches traveling towards the edges of the bosonic bath and then are 
reflected back to the trap center where they collide [Fig. \ref{fig:one_body_repul} (e)]. 
Consequently, when the two density fragments appear in $\rho^{(1)}_{\uparrow}(x;t)$ the impurities reside in two different two-body configurations 
[Figs. \ref{fig:two_body_repulsive} ($b_2$), ($b_4$) and ($b_6$)]. 
Namely the bosonic impurities either lie together at a certain density branch [see the diagonal elements of $\rho^{(2)}_{\uparrow\uparrow}(x_1,x_2;t)$] or they remain 
spatially separated with one of them residing in the left and the other in the right density branch [see the anti-diagonal elements of $\rho^{(2)}_{\uparrow\uparrow}(x_1,x_2;t)$]. 
Moreover, during their collision at $x=0$ the impurities are very close to each other as it is evident by the enhanced two-body probability 
in the neighborhood of $x_1=x_2=0$ [Fig. \ref{fig:two_body_repulsive} ($b_3$), ($b_5$)]. 
The dynamics of two weakly repulsive ($g_{II}=0.2$) impurities shows similar two-body correlation patterns to the non-interacting ones, 
as it can be seen by comparing Figs. \ref{fig:two_body_repulsive} ($b_1$)-($b_6$) to ($c_1$)-($c_6$). 
This behavior complements the similarities already found at the single-particle level (see sec. \ref{sec:density_repulsive}). 
The major difference on the two-body level between the $g_{II}=0.2$ and $g_{II}=0$ scenario is that in the former case 
$\rho^{(2)}_{\uparrow\uparrow}(x_1,x_2;t)$ is more elongated along its anti-diagonal when the impurities collide at $x=0$ [Figs. \ref{fig:two_body_repulsive} ($c_1$), ($c_3$)]. 
Therefore weakly interacting impurities
tend to be further apart compared to the $g_{II}=0$ case,
a result that reflects their direct repulsion. 
Other differences observed at the same time-instant in $\rho^{(2)}_{\uparrow\uparrow}(x_1,x_2;t)$ between the interacting and the non-interacting cases 
are due to the repulsive $s$-wave interaction that directly competes with the attractive induced interactions emanating in the system. 
For instance, shortly after a collision point e.g. at $t=55$, shown in Figs. \ref{fig:two_body_repulsive} ($b_5$) and ($c_5$), 
we observe that due to the repulsive $s$-wave interactions the attractive contribution between the impurities, see the diagonal 
of $\rho^{(2)}_{\uparrow\uparrow}(-2<x_1<2,-2<x_2<2;t)$ in Fig. \ref{fig:two_body_repulsive} ($b_5$) disappears [Fig. \ref{fig:two_body_repulsive} ($c_5$)]. 
\begin{figure*}[ht]
  	\includegraphics[width=0.7\textwidth]{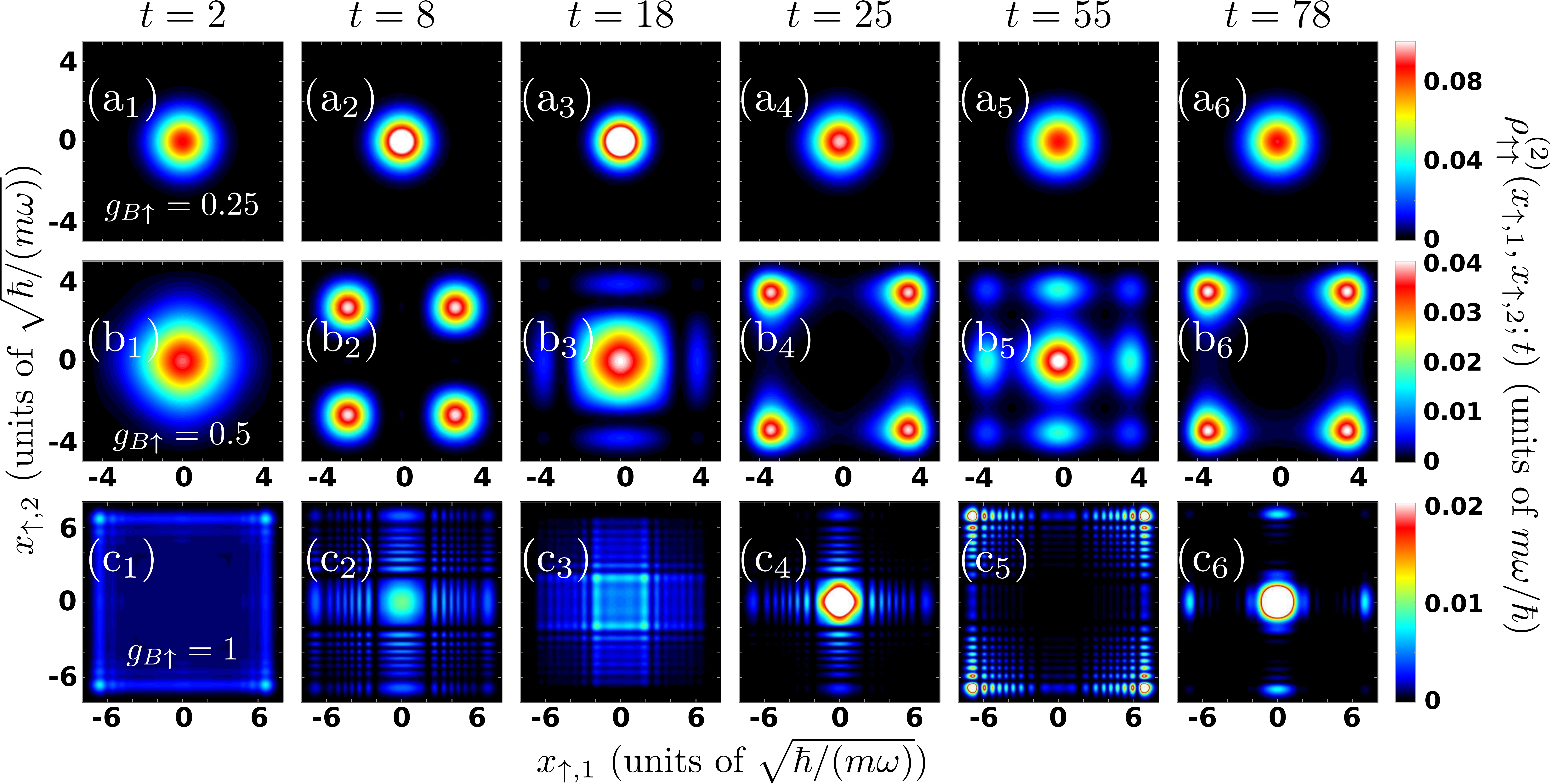}
  	\caption{Snapshots of the two-body reduced density 
  	matrix, $\rho^{(2)}_{\uparrow\uparrow}(x_1,x_2;t)$, of the two pseudospin-$\uparrow$ non-interacting 
  	($g_{II}=0$) bosonic impurities within the effective potential picture 
  	when considering an interspecies interaction quench to ($a_1$)-($a_6$) $g_{B\uparrow}=0.25$, 
  	($b_1$)-($b_6$) $g_{B\uparrow}=0.5$ and ($c_1$)-($c_6$) $g_{B\uparrow}=1.5$. 
  	The harmonically trapped bosonic mixture consists of $N_B=100$ atoms with $g_{BB}=0.5$ and 
  	$N_I=2$ impurities and it is prepared in its corresponding 
  	ground state configuration. }
  	\label{fig:two_body_repulsive_eff} 
\end{figure*}

Turning to very strong repulsions, e.g. for $g_{B\uparrow}=1.5$ lying in region $R_{IV}$, the correlation patterns of the two non-interacting 
impurities [Figs. \ref{fig:two_body_repulsive} ($d_1$)-($d_6$)] show completely different characteristics compared 
to the $g_{B\uparrow}\leq g_{BB}$ regime. 
Note here that in the dynamics of $\rho^{(1)}_{\uparrow}(x;t)$ the initially formed density hump breaks into two density fragments [Fig. \ref{fig:one_body_repul} (h)] 
possessing a multihump shape [see also Fig. \ref{fig:effective_pot_repulsive} (f)]. 
Subsequently, the fragments lying symmetrically with respect to $x=0$ perform a damped oscillatory motion in time residing around the edges of the Thomas-Fermi radius 
of the bosonic gas. 
The corresponding two-body reduced density matrix shows a pronounced probability peak around $x_1=x_2=0$ [Fig. \ref{fig:two_body_repulsive} ($d_1$)] indicating that 
at the initial stages of the dynamics the impurities are mainly placed together in this location. 
As time evolves, the impurities predominantly move as a pair towards the edge of the Thomas-Fermi background, see in particular the diagonal of 
$\rho^{(2)}_{\uparrow\uparrow}(x_1,x_2;t)$ in Figs. \ref{fig:two_body_repulsive} ($d_2$), ($d_3$), and simultaneously they start to exhibit 
a delocalized behavior as can be deduced by the small values of the off-diagonal elements of $\rho^{(2)}_{\uparrow\uparrow}(x_1,x_2\neq x_1;t)$. 
Entering deeper in the evolution the aforementioned delocalization of the impurities becomes more enhanced since $\rho^{(2)}_{\uparrow\uparrow}(x_1,x_2;t)$ 
disperses as illustrated in Figs. \ref{fig:two_body_repulsive} ($d_4$), ($d_5$) and ($d_6$). 
This dispersive behavior of $\rho^{(2)}_{\uparrow\uparrow}(x_1,x_2;t)$ is inherently related to the multihump structure of $\rho^{(1)}_{\uparrow}(x;t)$ and suggests 
from a two-body perspective the involvement of several excited states during the impurity dynamics. 
It is also worth mentioning that at specific time instants the diagonal of $\rho^{(2)}_{\uparrow\uparrow}(x_1,x_2;t)$ is predominantly populated 
[Figs. \ref{fig:two_body_repulsive} ($d_2$), ($d_3$), ($d_5$)] which is indicative of the presence of induced interactions.

\subsubsection{Two-body dynamics within the effective potential picture}\label{sec:two_body_repulsive_eff_pot_repul}

To further expose the necessity of taking into account the intra- and the interspecies correlations of the system in order to accurately describe 
the MB dynamics of the impurities we next solve the time-dependent Schr{\"o}dinger equation that governs the system's dynamics 
relying on the previously introduced effective potential picture [Eq. (\ref{effective_potential_repulsive})] via exact diagonalization \cite{ED_comment}. 
Thus our main aim here is to test the validity of $\bar{V}_I^{eff}(x)$  at least to qualitatively capture the basic features of the emergent 
nonequilibrium dynamics of the two impurities. 
We emphasize again that $\bar{V}_I^{eff}$ does not include any interspecies correlation effects that arise in the course of the temporal-evolution of the impurities. 
Within this approximation the effective Hamiltonian that captures the quench-induced dynamics of the impurities reads
\begin{eqnarray}
\begin{split}
H^{eff} = \int dx&~\hat{\Psi}^{\dagger}_{\uparrow} (x) \left( -\frac{\hbar^2}{2 m} \frac{d^2}{dx^2} + \bar{V}_I^{eff} \right) \hat{\Psi}_{\uparrow}(x)
\\& +g_{\uparrow \uparrow}\int dx \hat{\Psi}_{\uparrow}^{\dagger}(x)\hat{\Psi}_{\uparrow}^{\dagger}(x)\hat{\Psi}_{\uparrow}(x)\hat{\Psi}_{\uparrow}(x), 
\label{Heff_repul}
\end{split}
\end{eqnarray}
where $\hat{\Psi}_{\uparrow} (x)$ is the bosonic field-operator of the pseudospin-$\uparrow$ impurity and $g_{\uparrow \uparrow}$ denotes the 
intraspecies interactions between the two pseudospin-$\uparrow$ impurity atoms. 
Recall that the intercomponent contact interaction of strength $g_{B\uparrow}$ and the intraspecies interaction between the bath atoms are inherently 
embedded into $\bar{V}_I^{eff}$ [Eq. (\ref{effective_potential_repulsive})]. 
In particular, within $\bar{V}_I^{eff}$ we account for the correlated Thomas-Fermi profile of the BEC since $\rho^{(1)}_B(x;t)$ is determined from the MB approach. 
Below, we exemplarily study the dynamics of two non-interacting impurities and therefore we set $g_{\uparrow \uparrow}=0$ in Eq. (\ref{Heff_repul}). 
Moreover, in order to trigger the nonequilibrium dynamics we consider an interspecies interaction quench from $g_{B \uparrow}=0$ ($t=0$) 
to a finite repulsive value of $g_{B \uparrow}$. 
Such a sudden change is essentially taken into account via a deformation of $\bar{V}_I^{eff}$ [Eq. (\ref{effective_potential_repulsive})].  

The corresponding instantaneous two-body reduced density matrix of the impurities within $H^{eff}$ is depicted in Fig. \ref{fig:two_body_repulsive_eff} for 
distinct values of $g_{B\uparrow}$. 
Focusing on weak postquench interactions, e.g. $g_{B\uparrow}=0.25$, we observe that at the initial times the two-body dynamics of the impurities 
is adequately described within $H^{eff}$ [compare Figs. \ref{fig:two_body_repulsive} ($a_1$)-($a_3$) to Figs. \ref{fig:two_body_repulsive_eff} ($a_1$)-($a_3$)]. 
Indeed, in this time-interval only some minor deviations between the heights of the peaks of $\rho^{(2)}_{\uparrow\uparrow}(x_1,x_2;t)$ obtained within the MB 
and the $H^{eff}$ approach are observed. 
However, for longer times $H^{eff}$ [Fig. \ref{fig:two_body_repulsive_eff} ($a_4$)-($a_6$)] fails to capture the correct shape of $\rho^{(2)}_{\uparrow\uparrow}(x_1,x_2;t)$ 
and more precisely its deformations occuring along its diagonal or anti-diagonal [see Figs. \ref{fig:two_body_repulsive_eff} ($a_4$)-($a_6$)] which stem from the build up 
of higher-order correlations during the dynamics. 
\begin{figure*}[ht]
  	\includegraphics[width=0.9\textwidth]{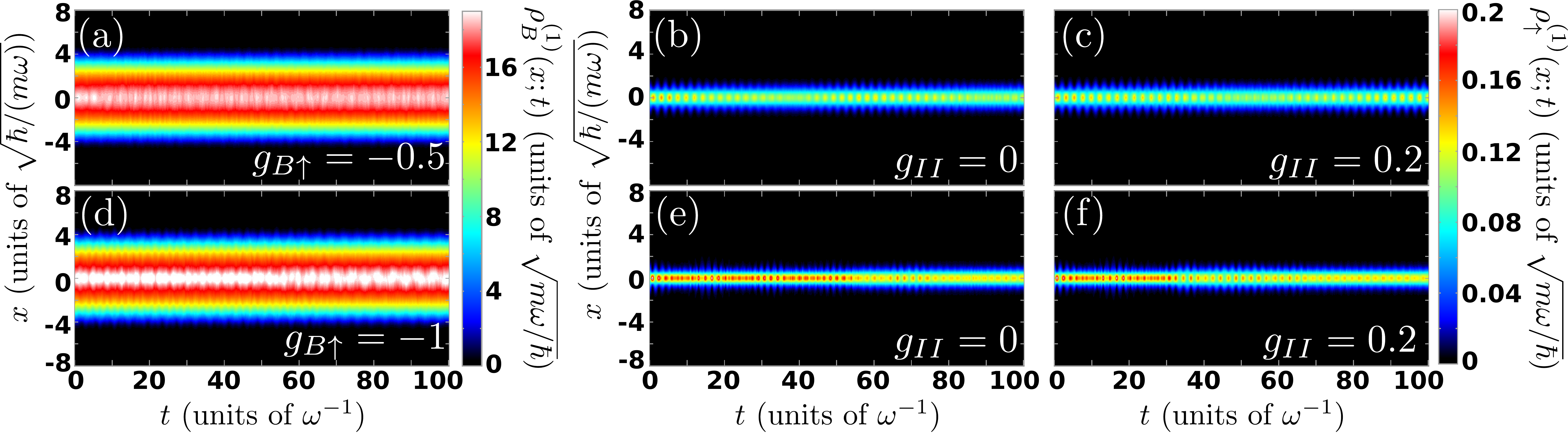}
  	\caption{Evolution of $\rho^{(1)}_{\sigma}(x;t)$ of (a), (d) the bosonic gas ($\sigma=B$), 
  	(b), (e) the pseudospin-$\uparrow$ part ($\sigma=\uparrow$) of the two non-interacting impurities, 
  	and that of (c), (f) two weakly interacting ($g_{II}=0.2$) impurities
  	for varying attractive postquench interspecies interaction strengths $g_{B\uparrow}$. 
  	In particular, in (a), (b), (c) $g_{B\uparrow}=-0.5$ and in (d), (e), (f) $g_{B\uparrow}=-1$.  
  	In all cases, the harmonically trapped bosonic mixture consists of $N_B=100$ bosons and $N_I=2$  
  	impurities with $g_{BB}=0.5$ and it is prepared in its corresponding ground state for $g_{B\uparrow}=0$.}
  	\label{fig:one_body_attract} 
\end{figure*} 

Increasing the repulsion such that $g_{B\uparrow}=0.5$, deviations 
between the effective potential approximation and the correlated approach become more severe. 
For instance, at the initial times the sharp two-body probability peak of $\rho^{(2)}_{\uparrow\uparrow}(x_1,x_2;t)$ in the vicinity of $x_1=x_2=0$ arising in the 
MB dynamics [Fig. \ref{fig:two_body_repulsive} ($b_1$)] becomes smoother within $H^{eff}$ [Fig. \ref{fig:two_body_repulsive_eff} ($b_1$)] 
although the overall shape of $\rho^{(2)}_{\uparrow\uparrow}(x_1,x_2;t)$ remains qualitatively similar. 
Moreover, the observed elongations along the diagonal of $\rho^{(2)}_{\uparrow\uparrow}(x_1,x_2;t)$ exhibited due to the presence of correlations are not 
captured in the effective picture, e.g. compare Figs. \ref{fig:two_body_repulsive} ($b_3$), ($b_5$) with Figs. \ref{fig:two_body_repulsive_eff} ($b_3$), ($b_5$). 
Remarkably, the two-body superposition identified in $\rho^{(2)}_{\uparrow\uparrow}(x_1,x_2;t)$ of two different two-body configurations occurring at 
specific time-instants is also predicted at least qualitatively via $H^{eff}$, see Figs. \ref{fig:two_body_repulsive_eff} ($b_2$), ($b_4$) and ($b_6$).   
We remark that the differences in the patterns of $\rho^{(2)}_{\uparrow\uparrow}(x_1,x_2;t)$ between $H^{eff}$ and the correlated approach are even more 
pronounced when $g_{II}=0.2$ (results not shown). 

Strikingly for strongly repulsive interactions, $g_{B\uparrow}=1.5$, $H^{eff}$ completely fails to capture the two-body dynamics of the impurities. 
This fact can be directly inferred by comparing $\rho^{(2)}_{\uparrow\uparrow}(x_1,x_2;t)$ within the two approaches, see Figs. \ref{fig:two_body_repulsive} ($c_1$)-($c_6$) and 
Figs. \ref{fig:two_body_repulsive_eff} ($c_1$)-($c_6$). 
Even at the initial stages of the dynamics the effective potential cannot adequately reproduce the correct shape of $\rho^{(2)}_{\uparrow\uparrow}(x_1,x_2;t)$, compare 
Fig. \ref{fig:two_body_repulsive_eff} ($c_1$) with Fig. \ref{fig:two_body_repulsive} ($d_1$). 
Note, for instance, the absence of the central two-body probability peak in the region $-2<x_1,x_2<2$ within $H^{eff}$ which demonstrates the  
correlated character of the dynamics. 
More precisely, $\rho^{(2)}_{\uparrow\uparrow}(x_1,x_2;t)$ obtained via $H^{eff}$ shows predominantly the development of two different two-body configurations. 
The first pattern suggests that the impurities either reside together at the same edge of the BEC background or each one is 
located at a distinct edge of the Thomas-Fermi profile, see e.g. Figs. \ref{fig:two_body_repulsive_eff} ($c_1$), ($c_5$). 
However, at different time-instants $\rho^{(2)}_{\uparrow\uparrow}(x_1,x_2;t)$ indicates that the impurities lie in the vicinity of the trap center 
as illustrated e.g. in Figs. \ref{fig:two_body_repulsive_eff} ($c_2$), ($c_4$) and ($c_6$), an event that never occurs for $t>5$ in 
the MB dynamics [see Fig. \ref{fig:one_body_repul} (h)]. 
It is also worth mentioning that the observed dispersive character of $\rho^{(2)}_{\uparrow\uparrow}(x_1,x_2;t)$ in the MB dynamics 
[see e.g. Fig. \ref{fig:two_body_repulsive} ($d_4$)-($d_6$)] is a pure correlation effect 
and a consequence of the participation of a multitude of excited states in the impurity dynamics which is never captured within $H^{eff}$.

\subsection{Quench to attractive interactions}\label{sec:attractive}

Next we discuss the dynamical behavior of both the BEC medium and the bosonic impurities on both the one- and the two-body level 
after an interspecies interaction quench from $g_{B\uparrow}=0$ to the attractive regime 
of $g_{B\uparrow}<0$. 
To explain basic characteristics of the dynamics of the impurities an effective potential picture is also employed. 
As in the previous section we first examine the emergent time-evolution of two non-interacting impurities ($g_{II}=0$) and then compare our findings to that 
of two weakly interacting ($g_{II}=0.2$) ones. 
\begin{figure*}[ht]
  	\includegraphics[width=0.9\textwidth]{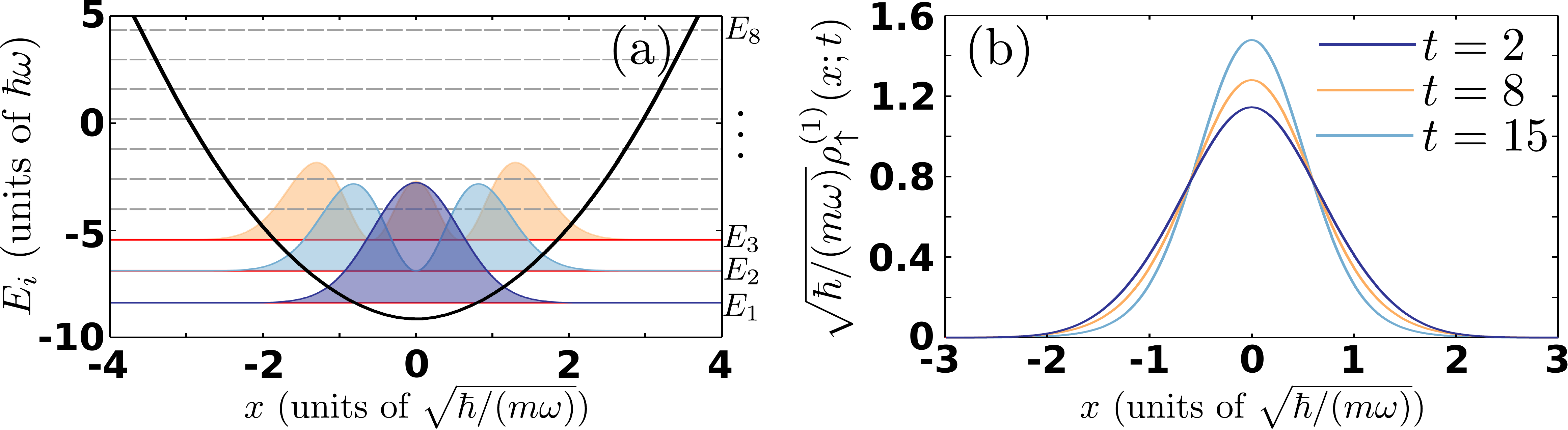}
  	\caption{Time-averaged effective potential, $\bar{V}_I^{eff}(x)$, over $T=100$ 
  	[Eq. (\ref{effective_potential_repulsive})] of the impurities for interspecies 
  	attractions $g_{B\uparrow}=-0.5$.  
  	The corresponding densities of the single-particle eigenstates and eigenenergies 
  	$E_i$, $i=1,2,\dots$ of $\bar{V}_I^{eff}(x)$ are also depicted. 
  	Instantaneous single-particle density profiles of the two non-interacting impurities for an 
  	interspecies interaction quench to 
  	$g_{B\uparrow}=-0.5$ within the MB approach.}
  	\label{fig:effect_pot_attract} 
\end{figure*}

\subsubsection{Single-particle dynamics and effective potential}\label{sec:density_attractive}

To investigate the spatially resolved dynamics of the multicomponent system after an interaction quench from $g_{B\uparrow}=0$ to $g_{B\uparrow}<0$, we first analyze the spatio-temporal evolution 
of the $\sigma$-species single-particle density $\rho_{\sigma}^{(1)}(x;t)$. 
The dynamical response of $\rho_{\sigma}^{(1)}(x;t)$ triggered by the quench is presented in Fig. \ref{fig:one_body_attract} for postquench interspecies attractions 
$g_{B\uparrow}=-0.5$ [Figs. \ref{fig:one_body_attract} (a), (b), (c)] and $g_{B\uparrow}=-1$ [Figs. \ref{fig:one_body_attract} (d), (e), (f)].   

Inspecting the dynamics of two non-interacting impurities at $g_{B\uparrow}=-0.5$ (region $R'_{III}$), shown in Figs. \ref{fig:one_body_attract} (a), (b), we deduce that 
$\rho_{\uparrow}^{(1)}(x;t)$ undergoes a breathing motion inside $\rho_{B}^{(1)}(x;t)$ characterized by a predominant frequency $\omega_{br}^{I}\approx 2.76$ and 
a secondary one $\omega_{br}^{'I}\approx 2.88$ thus producing a beating pattern. 
These two distinct frequencies stem from the center-of-mass and relative coordinate breathing modes of the impurities, whose existence 
originates from the presence of attractive induced interactions in the system. 
We remark that the breathing frequency of the center-of-mass can be estimated in terms of the corresponding effective potential of the impurities, 
see also Eq. (\ref{effective_pot_impurity_attractive}). 
In particular for $g_{B\uparrow}=-0.5$, $\omega_{br}^I=2\sqrt{2.06}\approx2.87$ 
(see also the comment in Ref. \cite{eff_pot_attr}) which is in very good agreement with $\omega_{br}^{'I}$. 
The relevant contraction of $\rho_{\uparrow}^{(1)}(x;t)$ can be inferred by its 
increasing amplitude that takes place from the very early stages of the nonequilibrium dynamics [Fig. \ref{fig:effect_pot_attract} (b)]. 
The beating pattern can be readily identified e.g. by comparing the maximum height of 
$\rho_{\uparrow}^{(1)}(x;t)$ during its contraction at initial and later stages of the dynamics, 
see e.g. $\rho_{\uparrow}^{(1)}(x;t)$ at $t=10$ and $t=40$ in Fig. \ref{fig:one_body_attract} (b). 
Moreover, as a consequence of the motion of the impurity and the relatively weak interspecies attraction, i.e. $g_{B\uparrow}=-0.5$, the Thomas-Fermi cloud of the bosonic gas 
becomes slightly distorted. 
In particular, a low amplitude density hump is imprinted on $\rho_{B}^{(1)}(x;t)$ exactly at the 
position of $\rho_{\uparrow}^{(1)}(x;t)$ as shown by the white colored region 
in Fig. \ref{fig:one_body_attract} (a) in the vicinity of $x=0$ \cite{Mistakidis_inject_imp}. 
An almost similar effect to the above-mentioned breathing dynamics is present also for the case of two weakly interacting impurities 
[Fig. \ref{fig:one_body_attract} (c)]. 
Here, the secondary frequency manifests itself at later evolution times resulting in turn
in a slower beating of $\rho_{\uparrow}^{(1)}(x;t)$ compared to the $g_{II}=0$ scenario 
[hardly visible in Fig. \ref{fig:one_body_attract} (c)]. 
This delayed occurrence is attributed to the presence of intraspecies repulsion which competes with 
the attractive induced interactions. 

For a larger negatively valued interspecies coupling, e.g. 
for $g_{BI}=-1$ within region $R'_{III}$, $\rho_{\uparrow}^{(1)}(x;t)$ becomes more spatially 
localized and again performs a decaying amplitude breathing motion, the so-called beating 
identified above, but with a larger major frequency, $\omega_{br}^{I}\approx3.2$, compared to the 
$g_{B\uparrow}=-0.5$ case [Fig. \ref{fig:one_body_attract} (e)]. 
Notice that the observed beating motion of the impurities persists while being more dramatic for 
this stronger attraction [compare Figs. \ref{fig:one_body_attract} (b) and (e)]. 
This enhanced attenuation of the breathing amplitude together with the strong localization of the 
impurities is a direct effect of the dominant presence of interspecies attractions between 
the impurity and the bath, see also Refs. \cite{Mistakidis_inject_imp}. 
Also, due to the stronger $g_{B\uparrow}$ and the increased spatial localization 
of $\rho_{\uparrow}^{(1)}(x;t)$, the density hump 
building upon $\rho_B^{(1)}(x;t)$ at the instantaneous position of the impurities is much more pronounced than that found for $g_{B\uparrow}=-0.5$ [Fig. \ref{fig:one_body_attract} (d)]. 
Note that the density hump appearing in $\rho_{B}^{(1)}(x;t)$ is essentially an imprint of the impurities presence and motion within the bosonic medium. 
Indeed, $\rho_{\uparrow}^{(1)}(x;t)$ exhibits a sech-like form tending to be more localized for a larger interspecies attractions $g_{B\uparrow}$, 
see e.g. $\rho_{\uparrow}^{(1)}(x;t)$ at a fixed time-instant for 
$g_{B\uparrow}=-0.5$ and $g_{B\uparrow}=-1$ in Figs. \ref{fig:one_body_attract} (b) and (e) respectively, a behavior that also holds 
for the consequent density hump in $\rho_{B}^{(1)}(x;t)$ [Figs. \ref{fig:one_body_attract} (a), (d)]. 
We should remark that for large negative $g_{B\uparrow}$ the system becomes strongly correlated and the BEC is highly excited. 
The latter is manifested by the development of an overall weak amplitude breathing motion of the bosonic gas, see Fig. \ref{fig:one_body_attract} (d). 
Furthermore, the inclusion of weak intraspecies repulsions between the impurities does not significantly alter their dynamics [Fig. \ref{fig:one_body_attract} (f)]. 
Indeed, a faint increase of their expansion magnitude takes place and the corresponding
amplitude of the beating decays faster [compare Figs. \ref{fig:one_body_attract} (d) and (f)]. 
\begin{figure*}[ht]
  	\includegraphics[width=0.7\textwidth]{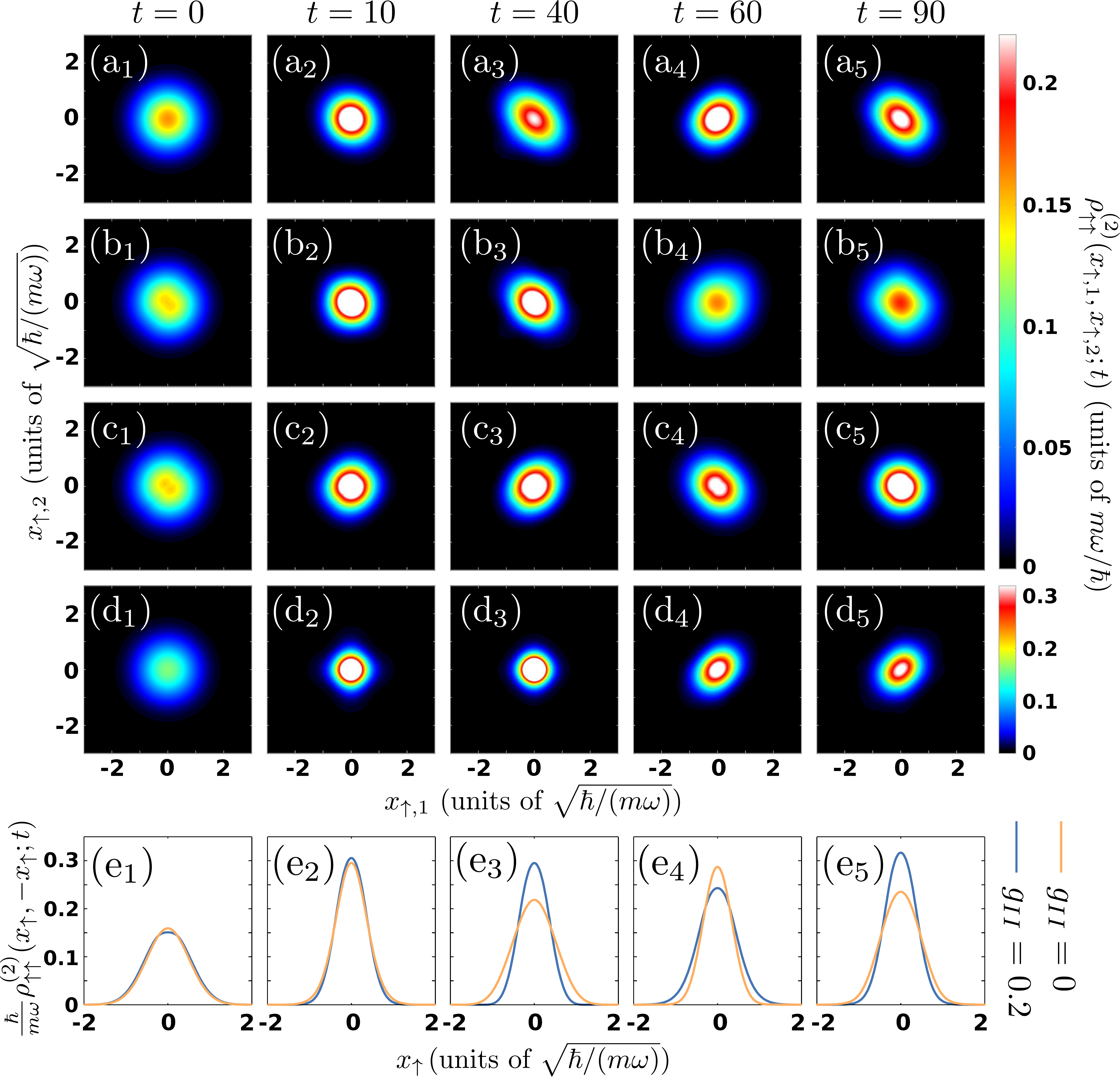}
  	\caption{Snapshots of $\rho^{(2)}_{\uparrow\uparrow}(x_1,x_2;t)$ (see legend), 
  	within the MB approach, of the two pseudospin-$\uparrow$ non-interacting ($g_{II}=0$) impurities  
  	upon considering an interaction quench from $g_{B\uparrow}=0$ to ($a_1$)-($a_5$) 
  	$g_{B\uparrow}=-0.5$ and ($d_1$)-($d_5$) $g_{B\uparrow}=-1$.   
  	($b_1$)-($b_5$) The same as in ($a_1$)-($a_5$) but for two weakly interacting ($g_{II}=0.2$) 
  	impurities in the correlated MB approach. 
  	($c_1$)-($c_5$) The same as in ($b_1$)-($b_5$) but within the effective potential 
  	approximation. 
  	(e$_1$)-(e$_5$) Instantaneous profiles of the antidiagonal of the two-reduced density $\rho^{(2)}_{\uparrow\uparrow}(x_{\uparrow},-x_{\uparrow};t)$ of two non-interacting 
  	[Figs. \ref{fig:two_body_attractive} ($a_1$)-($a_5$)] and two weakly interacting [Figs. \ref{fig:two_body_attractive} ($b_1$)-($b_5$)] impurities (see legend). 
  	The harmonically trapped Bose-Bose mixture is initially prepared in its corresponding ground 
  	state and consists of $N_B=100$ atoms with $g_{BB}=0.5$ 
  	and $N_I=2$ impurities.}
  	\label{fig:two_body_attractive} 
\end{figure*} 

The above-mentioned dynamics can also be qualitatively explained in terms of a 
corresponding effective potential approximation \cite{Mistakidis_orth_cat,Hannes,Mistakidis_inject_imp}. 
Yet again, the effective potential experienced by the impurities consists of the external harmonic oscillator $V(x)$ and the single-particle density 
of the BEC background. 
Importantly, since $ \rho_B^{(1)}(x;t)$ is greatly distorted from its original Thomas-Fermi profile due to the motion of the impurities, we invoke a time-averaged 
effective potential. 
Consequently, the effective potential of the impurity reads 
\begin{equation}
 \bar{V}_I^{eff}(x)=V(x)-\frac{\abs{g_{BI}}}{T} \int_0^T dt \rho_B^{(1)}(x;t), \label{effective_pot_impurity_attractive} 
\end{equation} 
where $T=100$ denotes the corresponding total propagation time. 
We remark that for the considered negative values of $g_{B\uparrow}$ the shape of $\bar{V}_I^{eff}(x)$ does not significantly change after averaging over $T=60$. 
A schematic illustration of $\bar{V}_I^{eff}(x)$ and the densities of its first few single-particle eigenstates at $g_{B\uparrow}=-1$ is presented 
in Fig. \ref{fig:effect_pot_attract} (a), see also remark \cite{eff_pot_attr}. 
The observed localization tendency of $\rho_{\uparrow}^{(1)}(x;t)$ around the aforementioned potential minimum is essentially determined by the 
strongly attractive behavior of $\bar{V}_I^{eff}(x)$. 
Remarkably, the distinct dynamical features of the impurities for an increasing interspecies attraction can be partly understood with the aid 
of $\bar{V}_I^{eff}(x)$. 
Indeed, for increasing $\abs{g_{BI}}$ the effective frequency of $\bar{V}_I^{eff}(x)$ is larger and $\bar{V}_I^{eff}(x)$ becomes more attractive. 
The former property of $\bar{V}_I^{eff}(x)$ accounts for the increasing breathing frequency of the impurity wavepacket for larger $\abs{g_{BI}}$. 
Additionally, the increasing attractiveness of $\bar{V}_I^{eff}(x)$ is responsible for the reduced width of $\rho_{\uparrow}^{(1)}(x;t)$ for a larger 
$\abs{g_{BI}}$ and thus its increasing localization tendency.

\subsubsection{Two-body correlation dynamics and comparison to the effective potential approximation}\label{sec:two_body_attractive}

Having described the time-evolution of the impurities on the single-particle level, we next analyze the dynamical response of the pseudospin-$\uparrow$ component 
by invoking the corresponding two-body reduced density matrix $\rho^{(2)}_{\uparrow\uparrow}(x_1,x_2;t)$ [see also Eq. (\ref{eq:two_body_density})]. 

The time-evolution of $\rho^{(2)}_{\uparrow\uparrow}(x_1,x_2;t)$ is depicted in Figs. \ref{fig:two_body_attractive} ($a_1$)-($a_5$) for two non-interacting ($g_{II}=0$) 
impurities following an interspecies interaction quench from $g_{B\uparrow}=0$ 
to $g_{B\uparrow}=-0.5$ (region $R'_{III}$). 
Before the quench the impurities lie together in the vicinity of the trap center since $\rho^{(2)}_{\uparrow\uparrow}(x_1=0,x_2=0;t=0)$ shows a high 
probability peak [Fig. \ref{fig:two_body_attractive} ($a_1$)]. 
However as time evolves the two bosons start to occupy a relatively smaller spatial region as can be deduced by the shrinking of the central two-body probability peak 
across the diagonal at $t=10$ 
in Fig. \ref{fig:two_body_attractive} ($a_2$). 
Then they move either opposite to each other [see the elongated anti-diagonal in Figs. \ref{fig:two_body_attractive} ($a_3$), ($a_5$)] or tend to bunch together at the same location 
[see the pronounced diagonal of $\rho^{(2)}_{\uparrow\uparrow}(x_1,x_2=x_1;t=60)$ in Fig. \ref{fig:two_body_attractive} ($a_4$)]. 
This latter behavior of the impurities is the two-body analogue of their wavepacket periodic expansion and contraction (relative coordinate breathing motion) discussed previously 
on the single-particle level [Fig. \ref{fig:one_body_attract} (b)]. 

The dynamics of two weakly repulsively interacting ($g_{II}=0.2$) impurities [Figs. \ref{fig:two_body_attractive} ($b_1$)-($b_5$)] shows similar characteristics 
to the above-described non-interacting scenario. 
Indeed, initially [Fig. \ref{fig:two_body_attractive} ($b_1$)] and at short times [Fig. \ref{fig:two_body_attractive} ($b_2$)] the impurities reside close to the trap 
center while later on they repel [see e.g. Fig. \ref{fig:two_body_attractive} ($b_3$)] or attract [Fig. \ref{fig:two_body_attractive} ($b_4$)] each other as a 
result of their breathing dynamics [see also Fig. \ref{fig:one_body_attract} (c)]. 
The major difference between the weakly interacting and the non-interacting impurities is that 
their distance which is given by the anti-diagonal distribution of 
their two-body reduced density matrix is slightly different, see Fig. \ref{fig:two_body_attractive} (e$_1$)-(e$_5$). 
For instance at $t=40$ the non-interacting impurities are further apart from each other as compared to the case of interacting impurities, while this situation is reversed at $t=90$. 
The aforementioned difference owes its existence to the distinct relative coordinate breathing frequencies. 
This can be directly inferred from the fact that $\rho^{(2)}_{\uparrow\uparrow}(x_1,x_2;t)$ possesses a larger spatial distribution when $g_{II}=0.2$ and it is attributed 
to their underlying mutual repulsion. 
For instance, even initially $\rho^{(2)}_{\uparrow\uparrow}(x_1,x_2;t=0)$ for $g_{II}=0.2$ [Fig. \ref{fig:two_body_attractive} ($b_1$)] is slightly deformed towards 
its anti-diagonal compared to the $g_{II}=0$ case [Fig. \ref{fig:two_body_attractive} ($a_1$)]. 
This behavior persists also during the evolution independently of the expansion or the contraction of the impurity cloud, as can be seen by comparing 
Figs. \ref{fig:two_body_attractive} ($b_4$) to ($a_4$) and Figs. \ref{fig:two_body_attractive} ($b_5$) to ($a_5$). 

To reveal the importance of both intra- and interspecies correlations for the impurity dynamics we then utilize the effective potential, $\bar{V}_I^{eff}(x)$, 
introduced in Eq. (\ref{effective_pot_impurity_attractive}) and solve numerically the time-dependent Schr{\"o}dinger equation of the impurities 
via exact diagonalization. 
We remark once more that $\bar{V}_I^{eff}$ neglects the interspecies correlations of the multicomponent system but includes the density profile of the BEC determined by the MB approach. 
In particular, we construct the effective Hamiltonian $H^{eff}$ of Eq. (\ref{Heff_repul}) but using the $\bar{V}_I^{eff}(x)$ of 
Eq. (\ref{effective_pot_impurity_attractive}). 
For brevity we focus on the case of $g_{\uparrow \uparrow}=0.2$ and analyze the dynamics after an interspecies interaction quench 
from $g_{B \uparrow}=0$ ($t=0$) to $g_{B \uparrow}=-0.5$. 
As explained in Sec. \ref{sec:two_body_repulsive_eff_pot_repul} within the effective potential picture this quench scenario accounts for the 
deformation of $\bar{V}_I^{eff}$. 
Snapshots of $\rho^{(2)}_{\uparrow\uparrow}(x_1,x_2;t)$ when $g_{II}=0.2$ and $g_{B\uparrow}=-0.5$ obtained within $H^{eff}$ are illustrated in 
Figs. \ref{fig:two_body_attractive} ($c_1$)-($c_5$). 
As it can be seen by comparing $\rho^{(2)}_{\uparrow\uparrow}(x_1,x_2;t)$ for the MB approach [Figs. \ref{fig:two_body_attractive} ($b_1$)-($b_5$)] 
and $H^{eff}$ [Figs. \ref{fig:two_body_attractive} ($c_1$)-($c_5$)] significant deviations occur between the two methods. 
Indeed, during the time-evolution the correlation patterns visible in $\rho^{(2)}_{\uparrow\uparrow}(x_1,x_2;t)$ calculated via $H^{eff}$ 
exhibit similar overall characteristics to the ones taking place in the correlated approach but at completely different time-scales. 
In fact, $\rho^{(2)}_{\uparrow\uparrow}(x_1,x_2;t)$ shows elongated shapes along its diagonal [Fig. \ref{fig:two_body_attractive} ($c_3$)] or 
anti-diagonal [Fig. \ref{fig:two_body_attractive} ($c_4$)] implying that the impurities tend to be relatively close or apart from one another 
respectively. 
The latter is again a manifestation of the breathing motion of the impurities at the two-body level. 
However $H^{eff}$ fails in general to adequately capture the correct spatial shape of $\rho^{(2)}_{\uparrow\uparrow}(x_1,x_2;t)$, 
since e.g. it predicts a repulsion of the impurities [Fig. \ref{fig:two_body_attractive} ($c_4$)] when in the presence of correlations they attract 
each other [Fig. \ref{fig:two_body_attractive} ($b_4$)] and vice versa [compare Figs. \ref{fig:two_body_attractive} ($c_3$) and ($b_3$)]. 
This difference is caused by the failure of the effective potential to account for induced interactions emanating within the MB setting. 

Finally, turning to strong postquench attractions within $R'_{III}$, e.g. for $g_{B\uparrow}=-1$ presented in Figs. \ref{fig:two_body_attractive} ($d_1$)-($d_5$), we observe that 
the two-body dynamics of the impurities is drastically altered with respect to the weakly attractive case $g_{B\uparrow}=-0.5$ described above. 
Initially, at $t=0$, the two bosons bunch together in the vicinity of the trap center since $\rho^{(2)}_{\uparrow\uparrow}(-1<x_1<1,-1<x_2<1;t=0)$ is predominantly 
populated [Fig. \ref{fig:two_body_attractive} ($d_1$)]. 
Subsequently the two-body distribution of $\rho^{(2)}_{\uparrow\uparrow}(x_1,x_2;t)$ spatially shrinks exhibiting a highly intense peaked structure 
around $-0.2<x_1,x_2<0.2$ as shown in Figs. \ref{fig:two_body_attractive} ($d_2$), ($d_3$). 
For longer evolution times $\rho^{(2)}_{\uparrow\uparrow}(x_1,x_2;t)$ deforms possessing an elongated shape across its diagonal 
[see Figs. \ref{fig:two_body_attractive} ($d_4$), ($d_5$)] which indicates that the impurities experience a mutual attraction. 
This latter behavior suggests the appearance of attractive induced interactions between the impurities mediated by the bosonic gas.

\section{Summary and Conclusions}\label{sec:conclusions}

We have investigated the ground state properties and the interspecies interaction quench quantum dynamics of two spinor bosonic impurities 
immersed in a harmonically trapped bosonic gas from zero to finite repulsive and attractive couplings. 
For two non-interacting impurities, we have shown that for an increasing attraction or repulsion their overall distance decreases 
indicating the presence of attractive induced interactions. 
Moreover, at strong attractions or repulsions the impurities acquire a fixed distance and bunch together either at the trap center or at the 
edge of the Thomas-Fermi profile of the bosonic gas respectively. 
For two weakly repulsive impurities we find that their ground state properties remain qualitatively the same for attractive couplings, but for repulsive 
interactions they move apart being located symmetrically with respect to the trap center. 
A similar to the above-described overall phenomenology takes place for smaller system sizes and heavier impurities. 

Regarding the quench dynamics of the multicomponent system we have analyzed the time-evolution of the contrast and its spectrum. 
We have revealed the emergence of six different dynamical response regions for varying postquench interaction strength which signify the existence, dynamical deformation and 
the orthogonality catastrophe of Bose polarons. 
We have also shown that the extent of these regions can be tuned via the intraspecies repulsion between the impurities, the impurity concentration and the size of the 
bath. 
Moreover, we have found that the polaron excitation spectrum depends strongly on the postquench interspecies interaction strength and the number of impurities 
but it is almost insensitive on the impurity-impurity interaction for the weak couplings. 

Focusing on weak postquench interspecies repulsions the non-interacting impurities perform a breathing motion manifested as a periodic expansion and contraction of their density on both 
the one- and two-body level. 
For an increasing repulsion the impurities single-particle density splits into two counterpropagating density branches that travel to the edges of the BEC medium 
where they are reflected back towards the trap center and subsequently collide, repeating this motion in a periodic manner. 
Here the impurities mainly reside in a superposition of two distinct two-body configurations, namely they either reside together or each one lies at a specific 
density branch, while during their collision they tend to remain very close to each other. 
In the strong repulsive regime we have observed that the density of the impurities shortly after the quench breaks into two fragments which are symmetric 
with respect to the origin and which exhibit a multihump structure and perform a damped oscillatory motion close to the Thomas-Fermi radius of the bosonic gas. 
This multihump structure leads to a spatially delocalized behavior of the corresponding two-body correlation patterns and suggests the involvement of 
higher excited states. 
In all cases the bosonic gas exhibits a breathing motion whose amplitude becomes more pronounced for an increasing repulsion.

Turning to attractive interspecies couplings, the impurities show a beating breathing motion and experience a spatial localization tendency at the trap center on both 
the one- and two-body level, a behavior that becomes more pronounced for larger attractions. 
Strikingly, for strong attractive interactions we unveil that gradually the impurities experience a mutual attraction on the two-body level. 
This effect demonstrates the pronounced presence of induced interactions for attractive interspecies ones. 
As a result of the impurities motion the density of the bosonic bath deforms, developing a low amplitude density hump located at the origin. 
The occurrence of this hump is a direct consequence 
of the presence of induced interactions. 

In all cases investigated in the present work, an intuitive understanding of the dynamics of the impurities is provided via an effective potential picture which is shown to be an adequate 
approximation for weak couplings where correlations are negligible. 
However, for increasing interaction strengths this effective model largely fails to adequately describe the dynamics on both the one- and two-body level due to the presence 
of both induced attraction and higher-order correlations. 
Finally, in all of the above-mentioned cases we showcase that a similar dynamical response takes place for two weakly repulsive impurities but the 
corresponding time-scales are slightly altered due to the competition between their mutual repulsion and the developed attractive induced interactions. 

There is a multitude of fruitful possible extensions of the present effort that can be addressed in future works. 
A intriguing aspect would be to examine whether thermalization of the impurities dynamics takes place for strong repulsions in the framework of the 
eigenstate thermalization hypothesis \cite{Rigol}. 
An imperative prospect is to study the robustness of the emergent quasiparticle picture in the current setting in the presence of temperature effects \cite{Tajima,Liu}. 
Moreover, the study of induced interactions of two bosonic impurities immersed in a Fermi sea would be an interesting prospect especially in order to expose their 
dependence on the different statistics of the medium. 
Additionally, the generalization of the present results to higher-dimensional settings would be highly desirable. 
Another interesting direction would be to investigate the collisional dynamics of subsonically or supersonically moving impurities in a lattice trapped bosonic gas.  
Here, one could unravel the properties of the emergent quasiparticles, such as their lifetime, residue, effective mass and induced interactions with respect to 
the interspecies interaction strength.

\appendix

\section{Remarks on the many-body simulations} \label{sec:convergence} 

To solve the underlying time-dependent MB Schr{\"o}dinger equation of the considered multicomponent system we invoke the 
Multi-Layer Multi-Configurational Time-Dependent Hartree Method for Atomic Mixtures (ML-MCTDHX) \cite{MLX,MLB1}. 
As discussed in Section \ref{sec:many_body_ansatz} it constitutes a variational approach for calculating the stationary and most importantly the 
nonequilibrium quantum dynamics of bosonic and fermionic multicomponent mixtures \cite{mistakidis_phase_sep,Mistakidis_eff_mass,Mistakidis_orth_cat} including 
spin degrees of freedom \cite{Mistakidis_orth_cat,Mistakidis_Fermi_pol,Koutentakis_prob}. 
A key advantage of the method is that it assumes the expansion of the total MB wavefunction in terms of a time-dependent and variationally 
optimized basis. 
Such a treatment enables us to capture both the intra- and intercomponent correlation effects by employing a computationally 
feasible basis size. 
The latter flexibility allows to span the relevant subspace of the Hilbert space efficiently for each time-instant which is in 
contrast to numerical methods relying on a time-independent basis. 
\begin{figure}[ht]
  	\includegraphics[width=0.46\textwidth]{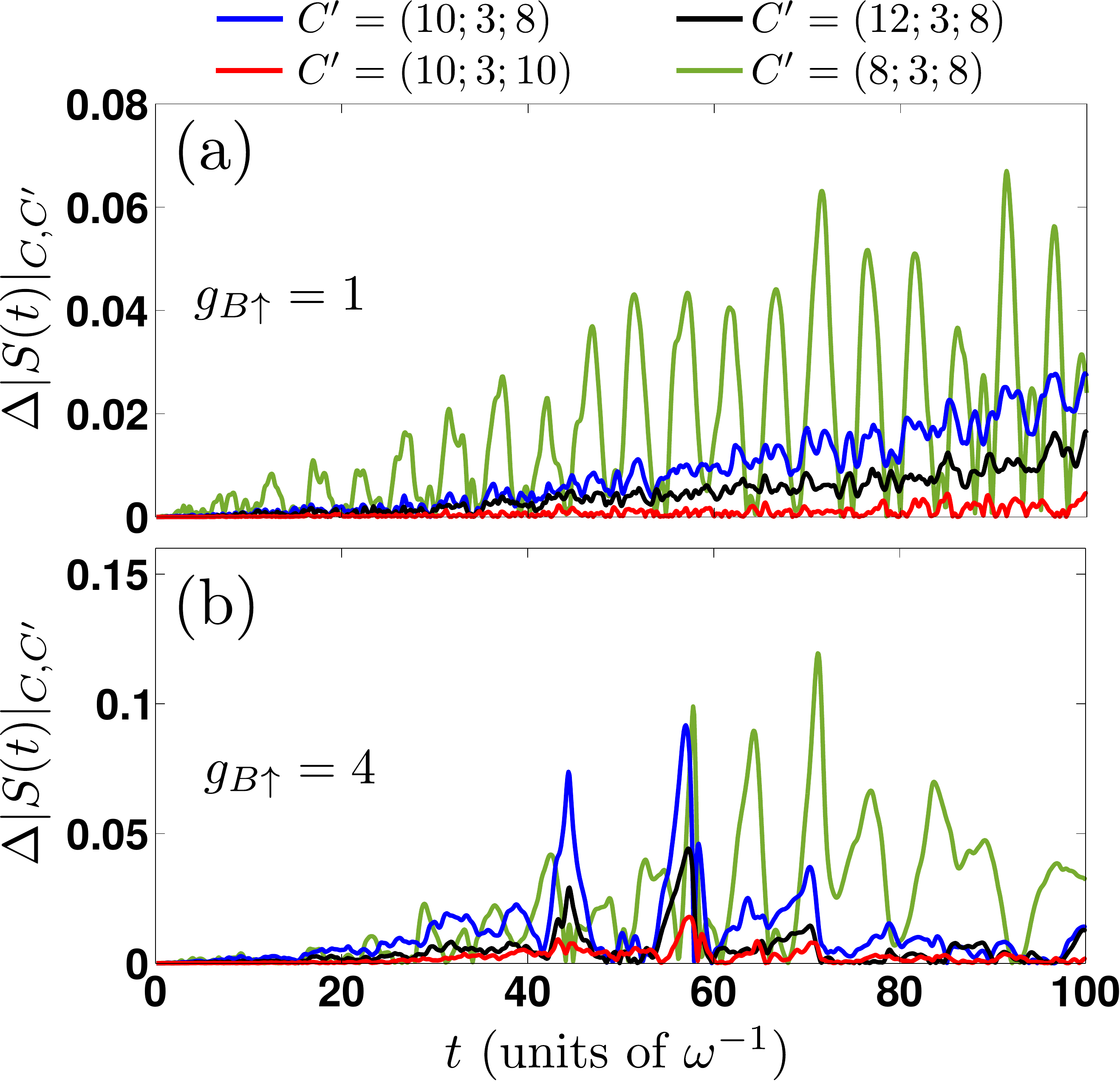}
  	\caption{ Temporal-evolution of the deviation of the two impurity contrast $\Delta\abs{S(t)}_{C,C'}$ between the $C=(12;3;10)$ and other 
  	orbital configurations $C'=(D;d^B;d^I)$ (see legend) for (a) $g_{B\uparrow}=1$ and (b) $g_{B\uparrow}=4$. 
  	In all cases $N_{B}=100$, $N_I=2$, $g_{BB}=0.5$ and $g_{II}=0$.}
  	\label{fig:convergence} 
\end{figure} 

The used Hilbert space truncation can be deduced from the employed orbital configuration space, denoted by $C=(D;d^B;d^I)$ 
with $D=D^B=D_I$ and $d^B$, $d^I$ being the number of species and single-particle functions of each species respectively 
[Eqs. (\ref{eq:wfn}), (\ref{eq:number_states}) and (\ref{eq:spfs})]. 
Additionally, within our implementation a sine discrete variable representation (sine-DVR) is utilized as the primitive basis 
for the spatial part of the SPFs with $\mathcal{M}=600$ grid points. 
The latter intrinsically introduces hard-wall boundary conditions at both edges of the numerical grid imposed herein at $x_\pm=\pm50$. 
We have ensured that the position of the hard-walls does not affect the presented results by assuring that no appreciable density occurs 
beyond $x_{\pm}=\pm20$. 
The eigenstates of the composite MB system are obtained by means of the so-called improved relaxation method \cite{MLX,MLB1} implemented in ML-MCTDHX. 
In order to simulate the nonequilibrium dynamics we propagate in time the wavefunction [Eq. (\ref{eq:wfn})] utilizing the appropriate Hamiltonian 
within the ML-MCTDHX equations of motion. 

To infer the convergence of our MB simulations we ensure that all observables of interest, e.g. $|\braket{\hat{\bm{S}}(t)}|$, $\rho_{\uparrow}^{(1)}(x;t)$, become to a certain degree insensitive 
upon varying the employed orbital configuration space chosen herein to be $C=(D;d^B;d^I)=(12;3;10)$. 
Below, we exemplarily showcase the convergence behavior of the contrast during evolution for a system composed of $N_B=100$ bosons 
with $g_{BB}=0.5$ and $N_I=2$ non-interacting ($g_{II}=0$) impurities. 
More precisely, we investigate its absolute deviation between the $C=(10;3;10)$ and other orbital configurations $C'=(D;d^B;d^I)$ during the 
nonequilibrium dynamics, namely   
\begin{equation}
\Delta\abs{S(t)}_{C,C'} =\frac{||\braket{\hat{\bm{S}}(t)}|_{C} -|\braket{\hat{\bm{S}}(t)}|_{C'}|}{|\braket{\hat{\bm{S}}(t)}|_{C}}. \label{converg_contrast} 
\end{equation} 

The time-evolution of $\Delta\abs{S(t)}_{C,C'}$ is illustrated in Fig. \ref{fig:convergence} after an interspecies interaction quench from 
$g_{B\uparrow}=0$ to intermediate repulsions e.g. $g_{B\uparrow}=1$ [Fig. \ref{fig:convergence} (a)] and strong ones such as $g_{B\uparrow}=4$ [Fig. \ref{fig:convergence} (b)]. 
As it can be readily seen by inspecting $\Delta\abs{S(t)}_{C,C'}$, a systematic convergence of $|\braket{\hat{\bm{S}}(t)}|$ can be achieved in both cases. 
At intermediate postquench repulsions, e.g. $g_{B\uparrow}=1$, $\Delta\abs{S(t)}_{C,C'}$ e.g. between the $C=(12;3;10)$ and $C'=(10;3;8)$ [$C'=(8;3;8)$] 
orbital configurations acquires a maximum value of the order of $3\%$ [$7\%$] at large propagation times as shown in Fig. \ref{fig:convergence} (a). 
As expected, an increasing $g_{B\uparrow}$ yields a larger relative error [Fig. \ref{fig:convergence} (b)] but still remaining at an adequately small 
degree. 
Indeed, turning to strong repulsions such as $g_{B\uparrow}=4$ we observe that the deviation $\Delta\abs{S(t)}_{C,C'}$ with $C=(12;3;10)$ 
and $C'=(12;3;8)$ [$C'=(10;3;8)$] lies below $5\%$ [$9\%$] throughout the evolution, see Fig. \ref{fig:convergence} (b). 
Finally, we should mention that a similar analysis has been performed for all other interspecies interaction strengths and observables discussed in the main text 
and found to be adequately converged (results not shown here for brevity).

\section*{Acknowledgements} 
S.I.M. and P.S. gratefully acknowledge financial support by the Deutsche Forschungsgemeinschaft 
(DFG) in the framework of the SFB 925 ``Light induced dynamics and control of correlated quantum
systems''. 
S. I. M  gratefully acknowledges financial support in the framework of the Lenz-Ising Award of the University of Hamburg.
T.B. has been supported by the Okinawa Institute of Science and Technology Graduate University.

{}


\begin{thebibliography}{60}
\bibitem{comment_few} Note that for $g_{B\uparrow}>g_{BB}=0.5$ the effective potential of Eq. (\ref{effective_potential_repulsive}) possesses 
a double-well structure as shown in Fig. \ref{fig:effective_pot_repulsive} (e). 
The width of its central barrier is determined by $R_{TF}$ which substantially decreases for smaller $N_B$. 
This decreasing tendency leads to a much more prominent overlap of the impurity wavefunction among the wells which in our case implies a smoother behavior 
of $\braket{r_{\uparrow\uparrow}}$.
\bibitem{comment_imbalance} Within the effective potential picture of Eq. (\ref{effective_potential_repulsive}) the miscibility/immiscibility transition 
is imprinted as a change in the shape of $\bar{V}_I^{eff}(x)$ from parabolic [Fig. \ref{fig:effective_pot_repulsive} (a)] to a 
double-well [Fig. \ref{fig:effective_pot_repulsive} (e)] potential. 
This transition occurs at $g_{B\uparrow}=\frac{m_I}{m_B}g_{BB}$ and 
therefore for $m_I>m_B$ is shifted to larger values of $g_{B\uparrow}$ than for $m_I=m_B$, a behavior that explains the shift 
of $\braket{r_{\uparrow\uparrow}}$ for heavy impurities.
\bibitem{ED_comment} Notice that the exact diagonalization simulations are performed within the two-body number state basis constructed by 
the single-particle states of a sine discrete variable representation consisting of 600 grid points, see also Appendix \ref{sec:convergence}.
\bibitem{eff_pot_attr}Notice here that the time-resolved form of the effective potential $V_I^{eff}(x,t)=V(x)-\abs{g_{BI}} \rho_B^{(1)}(x;t)$  
corresponds to a deformed attractive harmonic oscillator potential exhibiting a faint additional dip around $x\approx0$ 
resulting from the appearance of the density hump of $\rho_B^{(1)}(x;t)$ \cite{Mistakidis_inject_imp}. 
However in the averaged form of the effective potential this density dip contributes just as a shift of the frequency of the resulting 
parabolic potential. 
As an example at $g_{B\uparrow}=-0.5$ the effective trapping frequency $\omega^{eff}\approx \sqrt{2.06}$ within $\bar{V}_I^{eff}(x)$ 
while $\omega^{eff} = \sqrt{2}$ within $V_I^{eff}(x,t=0)$.
\end{thebibliography}
\end{document}